\documentclass[twoside,twocolumn,9pt]{article}
\usepackage{titling}

\usepackage{cuted}

\usepackage{extsizes}
\usepackage[super,sort&compress,comma]{natbib} 
\usepackage[version=3]{mhchem}
\usepackage[left=1.5cm, right=1.5cm, top=1.785cm, bottom=2.0cm]{geometry}
\usepackage{balance}
\usepackage{times,mathptmx}
\usepackage{sectsty}
\usepackage{graphicx} 
\usepackage{lastpage}
\usepackage[format=plain,justification=justified,singlelinecheck=false,font={stretch=1.125,small,sf},labelfont=bf,labelsep=space]{caption}
\usepackage{float}
\usepackage{fancyhdr}
\usepackage{fnpos}
\usepackage[english]{babel}
\addto{\captionsenglish}{%
  
}
\usepackage{array}
\usepackage{charter}
\usepackage[T1]{fontenc}
\usepackage[usenames,dvipsnames]{xcolor}
\usepackage{setspace}
\usepackage[compact]{titlesec}
\usepackage{hyperref}

\usepackage{epstopdf}

\definecolor{cream}{RGB}{222,217,201}

\usepackage{dcolumn}
\usepackage{bm}
\usepackage{color}

\DeclareMathOperator{\taninv}{Arctan}

\usepackage{authblk}

\newcommand{\beginsupplement}{%
        \setcounter{section}{0}
        \renewcommand{\thesection}{S\arabic{section}}%
        \setcounter{table}{0}
        \renewcommand{\thetable}{S\arabic{table}}%
        \setcounter{figure}{0}
        \renewcommand{\thefigure}{S\arabic{figure}}%
        \setcounter{equation}{0}
        \renewcommand{\theequation}{S\arabic{equation}}%
     }

\begin{document}

\pagestyle{fancy}
\thispagestyle{plain}
\fancypagestyle{plain}{

\renewcommand{\headrulewidth}{0pt}
}

\makeFNbottom
\makeatletter
\renewcommand\LARGE{\@setfontsize\LARGE{15pt}{17}}
\renewcommand\Large{\@setfontsize\Large{12pt}{14}}
\renewcommand\large{\@setfontsize\large{10pt}{12}}
\renewcommand\footnotesize{\@setfontsize\footnotesize{7pt}{10}}
\makeatother

\renewcommand{\thefootnote}{\fnsymbol{footnote}}
\renewcommand\footnoterule{\vspace*{1pt}%
\color{cream}\hrule width 3.5in height 0.4pt \color{black}\vspace*{5pt}} 

\makeatletter 
\renewcommand\@biblabel[1]{#1}            
\renewcommand\@makefntext[1]%
{\noindent\makebox[0pt][r]{\@thefnmark\,}#1}
\makeatother 
\renewcommand{\figurename}{\small{Fig.}~}
\sectionfont{\sffamily\Large}
\subsectionfont{\normalsize}
\subsubsectionfont{\bf}
\setstretch{1.125} 
\setlength{\skip\footins}{0.8cm}
\setlength{\footnotesep}{0.25cm}
\setlength{\jot}{10pt}
\titlespacing*{\section}{0pt}{4pt}{4pt}
\titlespacing*{\subsection}{0pt}{15pt}{1pt}

\fancyfoot{}
\fancyfoot[RO]{\footnotesize{\sffamily{\thepage}}}
\fancyfoot[LE]{\footnotesize{\sffamily{\thepage}}}
\fancyhead{}
\renewcommand{\headrulewidth}{0pt} 
\renewcommand{\footrulewidth}{0pt}
\setlength{\arrayrulewidth}{1pt}
\setlength{\columnsep}{6.5mm}
\setlength\bibsep{1pt}

\makeatletter 
\newlength{\figrulesep} 
\setlength{\figrulesep}{0.5\textfloatsep} 

\newcommand{\topfigrule}{\vspace*{-1pt}%
\noindent{\color{cream}\rule[-\figrulesep]{\columnwidth}{1.5pt}} }

\newcommand{\botfigrule}{\vspace*{-2pt}%
\noindent{\color{cream}\rule[\figrulesep]{\columnwidth}{1.5pt}} }

\newcommand{\dblfigrule}{\vspace*{-1pt}%
\noindent{\color{cream}\rule[-\figrulesep]{\textwidth}{1.5pt}} }

\makeatother

\twocolumn[
  \begin{@twocolumnfalse}
\vspace{3cm}
\sffamily
\begin{tabular}{m{4.5cm} p{13.5cm} }

 & \noindent\LARGE{\textbf{Predicting the conductance of strongly correlated molecules: the Kondo effect in perchlorotriphenylmethyl/Au junctions$^\dag$}} \\
\vspace{0.3cm} & \vspace{0.3cm} \\

 & \noindent\large {W. H. Appelt$^{a,b,\ddag}$, A. Droghetti$^{c,\ddag}$, L. Chioncel$^{b,d}$, M. M. Radonji\'c$^e$, E. Mu\~noz$^f$, \mbox{S. Kirchner}$^g$, D. Vollhardt$^{d}$, I. Rungger$^{h}$} \\

 & \noindent\normalsize{
Stable organic radicals integrated into molecular junctions represent a practical realization of the single-orbital Anderson impurity model. Motivated by recent experiments for perchlorotriphenylmethyl (PTM) molecules contacted to gold electrodes, we develop a method that combines density functional theory (DFT), quantum transport theory, numerical renormalization group (NRG) calculations and renormalized super-perturbation theory (rSPT) to compute both equilibrium and non-equilibrium properties of strongly correlated nanoscale systems at low temperatures effectively from first principles. We determine the possible atomic structures of the interfaces between the molecule and the electrodes, which allow us to estimate the Kondo temperature and the characteristic transport properties, which compare well with experiments. By using the non-equilibrium rSPT results we assess the range of validity of equilibrium DFT+NRG-based transmission calculations for the evaluation of the finite voltage conductance.
The results demonstrate that our method can provide qualitative insights into the properties of molecular junctions when the molecule-metal contacts are amorphous or generally ill-defined, and that it can further give a fully quantitative description when the experimental contact structures are well characterized.
} \\

\end{tabular}

 \end{@twocolumnfalse} \vspace{0.6cm}

  ]

\renewcommand*\rmdefault{bch}\normalfont\upshape
\rmfamily
\section*{}
\vspace{-1cm}

\footnotetext{$^{a}$Theoretical Physics II, Institute of Physics, University of
Augsburg, D-86135 Augsburg, Germany}
\footnotetext{$^{b}$Augsburg Center for Innovative Technologies, University of Augsburg, D-86135 Augsburg, Germany}
\footnotetext{$^{c}$Nano-Bio Spectroscopy Group and European Theoretical Spectroscopy Facility (ETSF), Centro de Fisica de Materiales, Universidad del Pais Vasco, Avenida Tolosa 72, 20018 San Sebastian, Spain}
\footnotetext{$^{d}$Theoretical Physics III, Center for Electronic
Correlations and Magnetism, Institute of Physics, University of
Augsburg, D-86135 Augsburg, Germany}
\footnotetext{$^{e}$Scientific Computing Laboratory, Center for the Study of Complex Systems, Institute of Physics Belgrade, University of Belgrade, Pregrevica 118, 11080 Belgrade, Serbia}
\footnotetext{$^{f}$Facultad de Fisica, Pontificia Universidad Cat\'olica de Chile, Casilla 306, Santiago 22, Chile}
\footnotetext{$^g$ Zhejiang Institute of Modern Physics, Zhejiang University, Hangzhou, Zhejiang 310027, China}
\footnotetext{$^h$ National Physical Laboratory, Teddington, TW11 0LW, United Kingdom; E-mail: ivan.rungger@npl.co.uk}

\footnotetext{\dag~Article published in the journal \textit{Nanoscale} (2018), DOI: 10.1039/c8nr03991g}

\footnotetext{\ddag These authors contributed equally to this work
\\
\vspace{0.cm}
\\
This article is licensed under a Creative Commons Attribution-NonCommercial 3.0 Unported License (CC)-BY-NC}



\section{\label{sec:intro}Introduction}

Molecular electronics holds great promise for future applications in computing, sensing, clean-energy, and even data-storage technologies~\cite{Aradhya,Bogani,Xiang}. However, a general difficulty so far has been the poor characterization of the device structures and their relationship with the measured conductances and functionalities. For this problem, {\it ab-initio} simulations based on density functional theory (DFT)~\cite{Kohn} have proven very successful in supporting experiments, and they have played a key role in advancing the field during the last decade~\cite{PhysRevB.63.245407,PhysRevB.65.165401,smeagol,PhysRevB.66.035322,Pecchia}. Yet, standard DFT-based transport schemes for simulations of experimental molecular junctions have several limitations. The most prominent of these is the failure to account for the strong electron correlations leading to the Kondo effect in devices comprising magnetic molecules, and rigorous treatments and extensions overcoming this problem are currently under active development~
\cite{Stefanucci, Troster, Bergfield,Stefanucci2,Kurth,Jacob_kurth}.

In this article, we establish a suitable combination of DFT and many-body techniques to achieve an unprecedented quantitative description of the equilibrium and non-equilibrium conductance of molecular devices showing Kondo effect. By using gold/perchlorotriphenylmethyl (PTM)/gold junctions as a specific example we relate the Kondo temperature to the electrode-molecule contact geometries, thus matching the range of variability of the experimental results~\cite{frisenda2015kondo}. Furthermore we address the dependence of the conductance at finite temperature and extend the method to finite bias.

Our multi-scale approach combines DFT, non-equilibrium Green's functions (NEGF)~\cite{Datta}, numerical renormalization group (NRG) methods~\cite{RevModPhys.47.773,PhysRevB.21.1003,PhysRevB.21.1044,RevModPhys.80.395}, and renormalized superperturbation theory (rSPT)~\cite{Munoz.11,Munoz.16}. First the contact geometry and electronic structure of molecular junctions are obtained by DFT+NEGF. Then the DFT Kohn-Sham (KS) states are projected onto an effective Anderson impurity model~\cite{Droghetti.17,Jacob1,Jacob2,Jacob3,Jacob4,Jacob5,Liviu,Requist,Lucignano,Baruselli1}, which is solved exactly to obtain the Kondo temperature and the equilibrium zero-temperature conductance via NRG. Based on these results we finally compute the non-equilibrium rSPT transport coefficients, which encode the behavior of the junctions at low temperature, finite magnetic field, and finite bias voltage~\cite{Munoz.11}. 

Stable organic radicals contacted to metal electrodes, such as the PTM molecule on Au, form a practical realization of the prototypical single-orbital Anderson impurity model~\cite{frisenda2015kondo,ji.hi.13,zh.ka.13,Requist}, and are therefore ideally suited to study the fundamental aspects of the interaction of magnetic impurities with metallic surfaces. These aspects include the interplay between the binding geometry and the energy level alignment with respect to the surface Fermi energy, as well as the electron correlations leading to the Kondo effect. 

\begin{figure}
	\centering
	\includegraphics[width=0.45\textwidth]{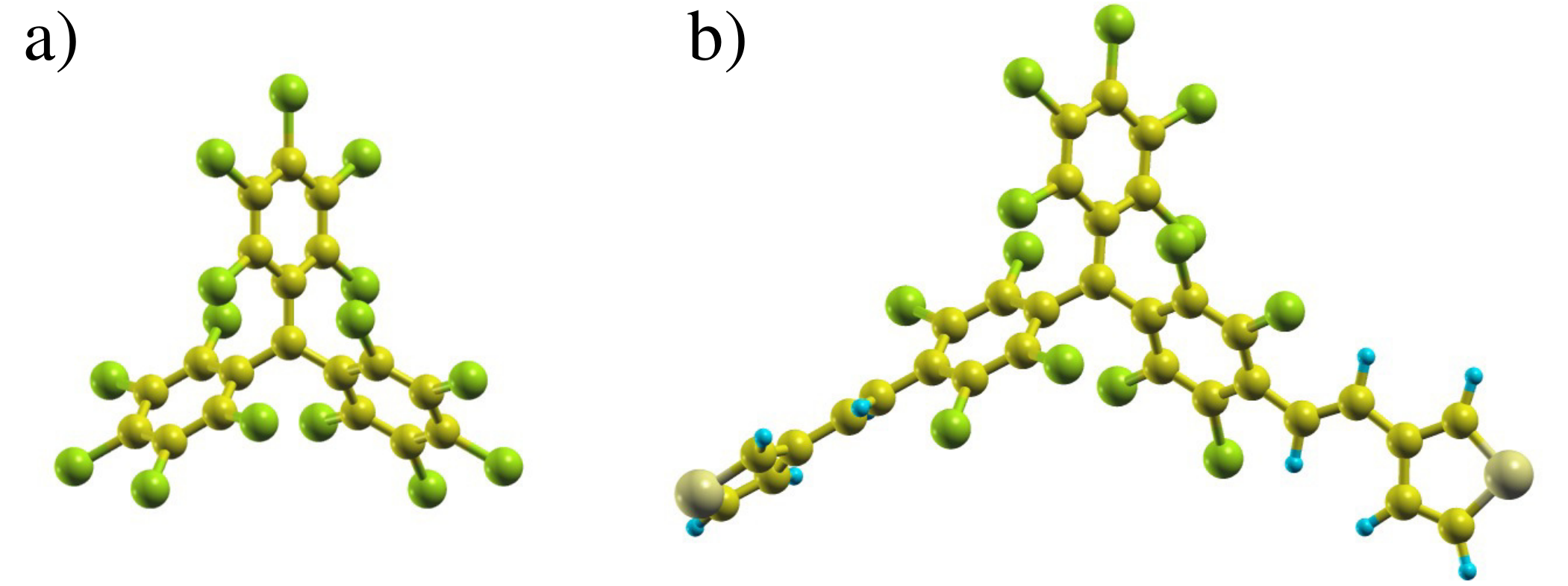}
\caption{Relaxed atomic structures of the bare PTM molecule (a) and of PTM-{\it bis}-thiophene (b) (green spheres represent Cl atoms, blue spheres H, large yellow spereres S, and smaller dark yellow spheres represent C).}
\label{fig:fig0}
\end{figure}

In recent experiments~\cite{frisenda2015kondo,Francesc} PTM-radicals were functionalized with thiophene linkers producing the PTM-{\it bis}-thiophene radical (called PTM-BT in the following to distinguish it from the bare PTM; see also Fig. \ref{fig:fig0} for their atomic structures). 
These molecules were then integrated into gold mechanically-controlled break-junctions (MCBJs) and gold electromigrated break-junctions (EMBJs) to measure their transport properties. While at room temperature very low conductance values were reported~\cite{Francesc}, at low temperature a zero-bias conductance resonance was observed in many of the junctions, and its Kondo character verified by temperature- and magnetic field-dependent measurements~\cite{frisenda2015kondo}. 
The low-temperature results indicate that the PTM radical can preserve the unpaired spin in a solid state three-terminal configuration, and that it is stable under mechanical stretching of the electrodes. 
One of the remarkable features is the rather high Kondo temperature of about 3 K, which is largely constant upon stretching of the junction. 
This implies that for the junctions that exhibiting Kondo behavior the contact of the molecule to one of the electrodes is very strong, and is not affected by the elongation of the junction in the MCBJ process. 
In contrast, the background conductance shows large variations. This can happen upon stretching when the contact to the second electrode varies significantly, or else when one of the two electrodes changes its Au-Au bond conformation significantly~\cite{French,French2}. Overall the low-temperature experimental results point to a structure with highly asymmetric coupling to the electrodes. In the following we will show that this hypothesis is indeed confirmed by our calculations, thus providing a detailed understanding of the electronic and transport properties of the PTM/gold junctions at the atomic scale.  

The paper is organized as follows. We first discuss the equilibrium DFT results for a number of possible junction structures (Sec. \ref{sec.dft}), and then provide estimates for the Kondo temperature for these geometries (Sec. \ref{sec:NRG}). For a set of geometries we then present the linear response transport properties including the strong electron-electron correlations obtained by DFT+NEGF+NRG (Sec. \ref{subsec:trans}) and finally extend the results to finite temperature and finite bias via rSPT (Sec. \ref{sec:super-p}). 

\section{DFT calculations}
\label{sec.dft}
\begin{figure*}
	\centering
	\includegraphics[width=0.95\textwidth]{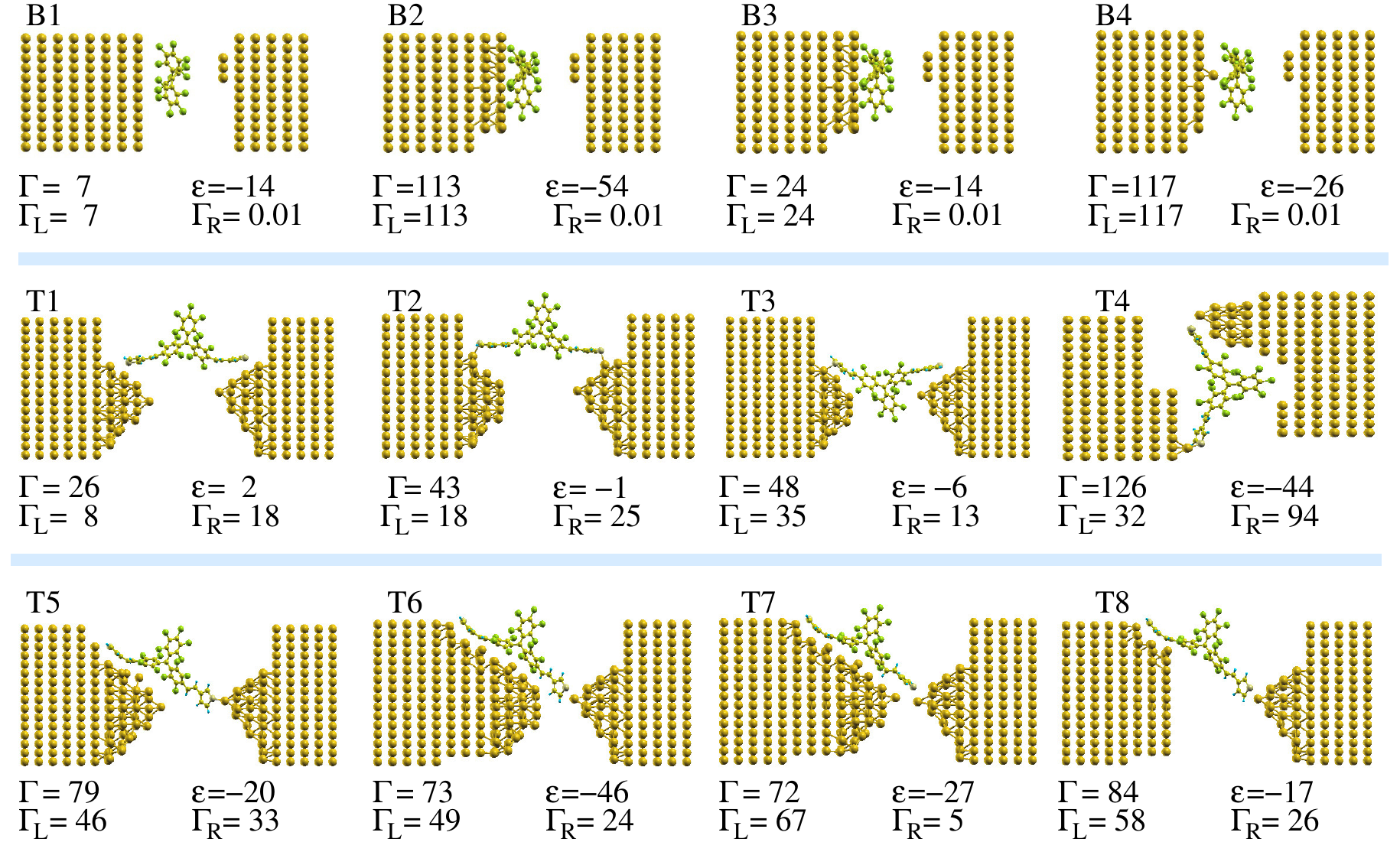}
\caption{Junction geometries for bare-PTM on an Au surface (B1-B4), and for PTM-BT between two Au electrodes (T1-T8), investigated in this paper. For each junction we specify the broadening of the singly occupied molecular orbital induced by the coupling to the electrodes ($\Gamma$), its position with respect to $E_\mathrm{F}$  ($\varepsilon$), and the coupling to the left and right electrodes ($\Gamma_\mathrm{L}$ and $\Gamma_\mathrm{R}$, respectively; $\Gamma=\Gamma_\mathrm{L}+\Gamma_\mathrm{R}$). All units of the specified quantities are meV.}
	\label{fig:fig5}
\end{figure*}

PTM has a propeller-like structure with a central carbon atom coordinated by the three phenyl rings. 
In the gas phase, it has the typical electronic structure of a radical~\cite{Nuria,Nuria2,Gonca}. The energy spectrum has doubly occupied electronic states filled up to the highest occupied molecular orbital (HOMO). Above the HOMO there is a further well-separated, singly occupied molecular orbital (SOMO) with an unpaired electron, giving a total molecular spin quantum number of 1/2. In the PTM, the charge isosurface indicates that the SOMO is mainly confined to the central carbon, 
while the HOMO and the lowest unoccupied molecular orbital (LUMO) are largely located on the rest of the molecule. This is presented more extensively in the ESI Sec. S2,
while the computational details of our DFT calculations are given in the ESI Sec. S1.
The difference between the ionization potential and electron affinity of the molecule defines the fundamental gap and
corresponds to the charging energy $U$. 
In the absence of any experimental results, we calculate $U$ 
via total energy differences
\cite{DroghettiCDFT}
to be about
$4$ eV. PTM-BT has a very similar electronic structure to that of the bare PTM, although the SOMO is slightly delocalized over the thiophene ligands~\cite{frisenda2015kondo}, and this results in a charging energy smaller by about $0.4$ eV.
Note that when the molecule is placed between Au electrodes there is a significant renormalization of the energy levels and consequently a reduction of the charging energy, which we discuss in Sec. S2 of the ESI as well as in Sec. \ref{sec:tkest}.

In order to understand the electronic structure of the molecule/Au contact and how this determines the key parameters affecting the Kondo temperature, we consider a number of qualitatively different model structures, which are shown in Fig.\ref{fig:fig5}. 
To start, we look at the ideal case of a bare PTM molecule on a flat Au(111) surface, which we denote as configuration (CFG) B1 in Fig.~\ref{fig:fig5}. 
The 3-atom Au tip is placed at a rather large distance, so that the electronic coupling 
between the molecule and the tip is negligible with respect to that to the substrate. 
Since in MCBJ and EMBJ experiments the Au stretched surface is expected to be highly corrugated rather than perfectly flat \cite{French,French2}, 
we then model a rough Au surface by removing a number of Au atoms from the perfect Au(111) surface (CFGs B2 to B4). 
Finally, we consider a number of break-junction setups comprising PTM-BT (CFGs T1 to T8). 
The detailed contact structure is expected to be different for each individual experimental conductance trace measurement. The model junctions considered here include cases with both symmetric and asymmetric molecule-electrodes coupling. 
For some structures the PTM central core is located inside the junction's empty gap, whereas for other structures it is physisorbed on one of the electrodes. 
Furthermore, the thiophene linkers can be connected to the electrodes either non-covalently or covalently via a sulfur-Au adatom direct bond.

\begin{figure}
	\centering
	\includegraphics[width=0.45\textwidth,clip=true]{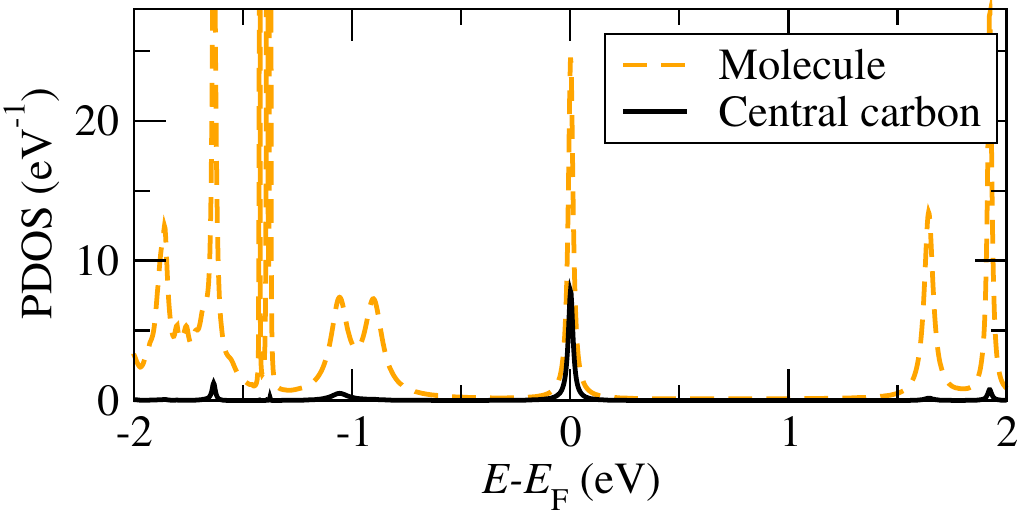}
\caption{LDA projected density of states (PDOS) for  the configuration $\mathrm{T1}$. The peak at the Fermi energy corresponds to the singly occupied molecular orbital, which defines our Anderson impurity, and is located mainly on the central carbon atom of the PTM.}
	\label{fig:fig6}
\end{figure}

A representative DFT projected density of states (PDOS) is shown in Fig. \ref{fig:fig6} (see ESI Sec. S1 for the computational details). When the molecule is in contact with the Au electrodes, the SOMO DOS
can be modeled approximately by a half-filled Lorentzian-like peak close to the Fermi energy, $E_\mathrm{F}$.
Note that while we refer to the state as SOMO also when the molecule is on the Au substrate for consistency, its occupation can generally deviate from one in this case.
The full width at half maximum (FWHM) of the SOMO peak corresponds to its electronic coupling to the Au substrate, $\Gamma$~\cite{Droghetti.17}, which
can be calculated by using the projection scheme recently developed in Ref.~\citenum{Droghetti.17}. The results for each model geometry considered are indicated in Fig.~\ref{fig:fig5} along with the DFT SOMO on-site energy, $\epsilon$, relative to $E_\mathrm{F}$. These values are the parameters required for the evaluation of the Kondo temperature. 

As a matter of notation we label the structures with the propeller-like PTM parallel (perpendicular) to the surface, 
as ``parallel'' (``perpendicular'') configurations.
For the idealized case of a bare PTM on a flat Au(111) surface, we find that the molecule is physisorbed with an energy difference between the ``parallel'' configuration (B1) and the ``perpendicular'' configuration (not shown) of about $335$ meV, favoring the ``parallel'' configuration. 
The equilibrium position of the central C atom is located at about $5.18$~\AA~ from the top Au layer. For this configuration there is a negligible charge-transfer from the surface to the molecule, and the PTM preserves its unpaired electron, with $\Gamma \approx 7$ meV and therefore very small.

On the corrugated surface (CFG B2) the molecule can bind better to the Au, since part of its phenyl rings can move into regions where the Au surface has a dip. 
In CFG B2 an Au atom is located below the central C atom of the bare-PTM. 
This atom is then removed in the CFG B3, while it is kept as the only atom from the top-most Au surface in the CFG B4. 
Comparing the $\Gamma$-values for these structures allows us to estimate the effect of Au atoms directly in contact with the central C atom of the PTM. 
For CFG B2 we find the occupation of the SOMO to be $1.40$ electrons, indicating that a partial electron transfer between the gold and the molecule has occurred. 
In fact, the SOMO DOS peak lies below $E_\mathrm{F}$ ($\epsilon=-54$ meV). 
The increased charge transfer indicates an increased screening of the transferred electrons by the Au surface atoms, which is due to the molecule moving closer to the Au surface,
in particular to the Au atom closest to the core of the PTM molecule. In general an increase in the screening also leads to a reduction of $U$ (see the discussion in the ESI Sec. S2).
We note that the results for the charge transfer
obtained for non-spin-polarized calcualtions are approximately the same as those obtained in spin-polarized calculations in the ESI Sec.~S2.
The electronic coupling of 114 meV for CFG B2 is much larger than the one for the PTM on flat Au(111). An analysis of the origin of such a large coupling shows that it is mainly due to the Au atom underneath the central C atom of the PTM. 
In fact, for CFG B3, where this central Au adatom is removed, the coupling drops to 23 meV, while it remains large for CFG B4, where only this Au adatom is kept of the top Au surface layer.

Finally, we consider the model break-junctions (CFGs T1 to T8). In these cases, we use only the ``perpendicular'' configuration, since the ``parallel'' PTM-BT configuration would require very large simulation cells, which are beyond our current computing resources. 
The results allow us to infer the general trends for the electronic coupling of the radical center to the Au electrodes through the thiophene linkers (see Fig. \ref{fig:fig5}). 
As can be seen the computed values of $\Gamma$ vary over almost one order of magnitude, from $26$ to $126$ meV. We note that in break-junctions $\Gamma$ is the sum of two contributions, $\Gamma_\mathrm{L}$ and $\Gamma_\mathrm{R}$, representing the electronic coupling to the left and right lead, which we calculate individually with the method outlined in Ref. \citenum{Droghetti.17}. 
In general we find that $\Gamma_\mathrm{L}$ or $\Gamma_\mathrm{R}$ are large when there are Au atoms close to the thiophene linkers, such as for $\Gamma_\mathrm{L}$ in CFG $\mathrm{T7}$.
A bond between the sulfur atoms and a protruding Au atom also increases the coupling. 
On the other hand, the coupling is low when such a bond is absent, and when the angle between the thiophene and the Au is larger, such as for $\Gamma_\mathrm{L}$ in CFG $\mathrm{T1}$.

\section{Kondo effect}
\label{sec:NRG}

\subsection{Formulation of the Single Impurity Anderson Problem}
\label{sec:NRG_SIAM}
The PTM in contact with the leads is modeled by a single impurity Anderson Model (SIAM), which has the Hamiltonian~\cite{PhysRev.124.41} 
 \begin{eqnarray}
H_\mathrm{SIAM} &=& H_d+H_c+H_\mathrm{hyb}, \label{siam}\\
H_d &=&\sum_\sigma \epsilon_d n_{d\sigma} + U  n_{d\uparrow} n_{d\downarrow}, \nonumber \\
H_c &=& \sum_{k,\sigma} \epsilon_{k } n_{k,\sigma}, \nonumber \\
H_\mathrm{hyb} &=& \sum_{k,\sigma} V_{k} (d_\sigma^\dagger c_{k,\sigma} +  c_{k,\sigma}^\dagger d_\sigma  ),\nonumber
\end{eqnarray}
where $H_d$ describes the electrons of spin $\sigma$ localized at the impurity site, which are created (annihilated) by the operator $d_\sigma^\dagger$ ($d_\sigma$), with
 $n_{d\sigma}=d^\dagger_{\sigma}d_{\sigma}$ being the corresponding number operator; $\epsilon_d$ is the orbital energy, $U$ the charging energy, and $\langle n_d\rangle=\sum_\sigma\langle d_\sigma^\dagger d_\sigma\rangle$ the occupation, where the bracket $\langle \dots \rangle$ denotes the thermal expectation value.
For the PTM molecule the impurity site is the SOMO. Note that $\epsilon_d$ does not coincide with $\epsilon$ in Fig. \ref{fig:fig6}, since the on-site Coulomb interaction is already partially accounted for in KS-DFT, and this contribution has to be subtracted, so that $\epsilon_\mathrm{d}=\epsilon-\epsilon_\mathrm{dc}$~\cite{Kotliar,Droghetti.17}. Here $\epsilon_\mathrm{dc}$ is the so-called double counting correction, whose exact expression is not known except for certain limiting cases, and several approximations have been introduced in the literature~\cite{pe.ma.03}.
In general $\epsilon_\mathrm{dc}$ depends on $U$, and in the commonly used ``fully localized limit'' it is given by $\epsilon_\mathrm{dc}=U (n_{d}^{\mathrm{DFT}}-1/2)$\cite{Bergfield}, where $n_{d}^{\mathrm{DFT}}$ is the DFT occupation of the impurity.
A comprehensive discussion of the difficulties arising when combining DFT with such an Anderson impurity model and more generally the dynamical mean field theory is given in References \cite{Katsnelson.2008,pe.ma.03}.
Note that instead of the Anderson impurity model one can also use other methods to treat the highly correlated subsystem, such as for example embedded correlated wavefunction schemes \cite{Carter.06,Carter.14}. A review of the advantages and limitations of various embedding schemes that link many-body calculations for a subsystem to an environment treated at the DFT level is given in Reference \cite{ChanGKL.16}.

Since in an experimental setting the occupation can be set by applying a gate voltage, here we treat $\epsilon_d$ as an adjustable parameter, independent of the DFT results, and choose its value to ensure a specified occupation of the impurity orbital.
We will also investigate how the results depend on the charging energy $U$, and will provide estimates of possible values of $U$ for PTM/Au geometries.

In Eq.~(\ref{siam}) $H_c$ describes the effective bath of electrons 
with momentum $k$ and spin $\sigma$, which are created (annihilated) by the operator $c^\dagger_{k,\sigma}$ ($c_{k,\sigma}$) and with number operator 
$n_{k,\sigma}=c^\dagger_{k,\sigma}c_{k,\sigma}$. The effective bath includes the electrons in the Au leads, as well as those in the molecular orbitals, except for the SOMO. 
We note that the chemical potential in the Hamiltonian Eq.~(\ref{siam}) is set to zero by shifting both the bath and impurity energies $\epsilon_d$ and $\epsilon_k$ by an additive constant.
This does not affect the properties of the system. Furthermore, in  zero-temperature calculations we will refer to the chemical potential $\mu=0$ as the Fermi energy $E_\mathrm{F}=0$, which is most commonly used in first principles calculations.

Finally, $H_\mathrm{hyb}$ accounts for the hybridization between the bath and the impurity,
with $V_k$ corresponding to the hybridization matrix element.
Accordingly, we can  define the hybridization function $\Delta(E)=\mathrm{Re}\Delta(E)+i\;\mathrm{Im}\Delta(E)$, with 
\begin{eqnarray}
\mathrm{Im}\Delta(E)&=&-\pi\sum_k |V_k|^2\delta(E-\epsilon_k),\label{imdelta}\\
\mathrm{Re}\Delta(E)&=&\frac 1 \pi \textrm{P} \int dE' \frac{\mathrm{Im}\Delta(E')}{E'-E},
\label{redelta}
\end{eqnarray}
and the coupling strength $\Gamma(E) =-2\;\mathrm{Im}\Delta(E)$. The DFT results for $\Gamma(E_\mathrm{F}=0)$ for several PTM/Au contacts are presented in the previous section, and the results are shown in Fig.~\ref{fig:fig5}.

\subsection{Estimation of the Kondo temperature}
\label{sec:tkest}

\begin{figure}
	\centering
	\includegraphics[width=0.47\textwidth,clip=true]{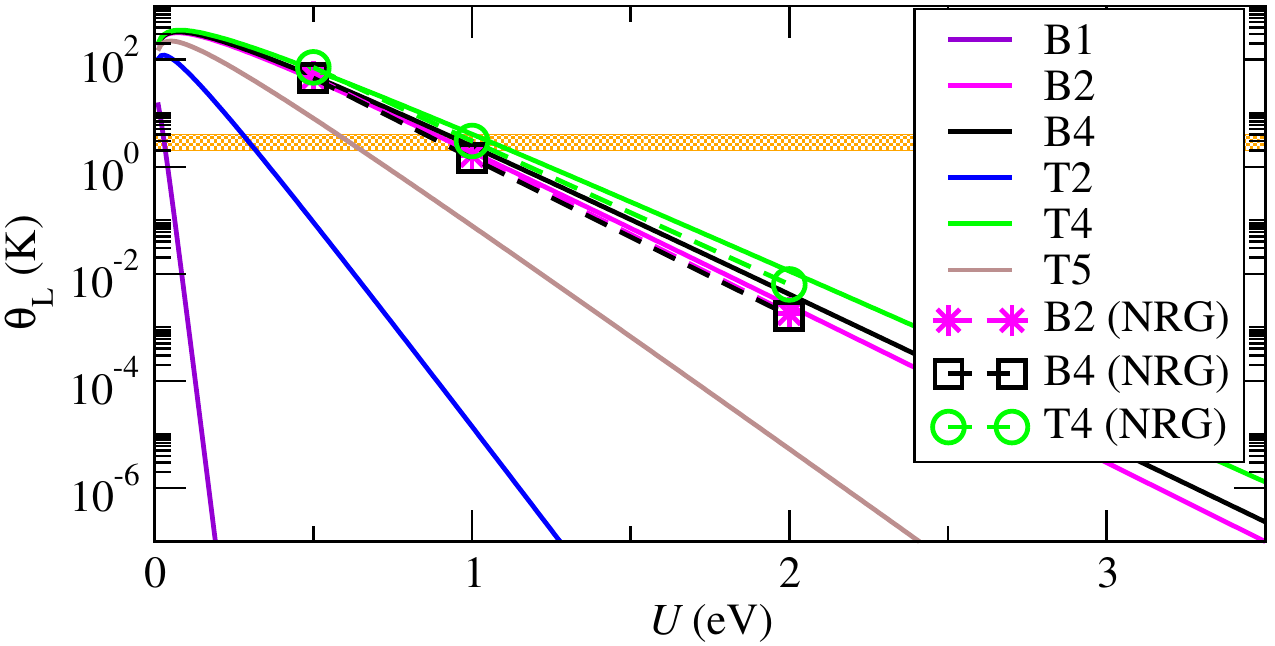}
\caption{ Kondo temperature, $\theta_\mathrm{L}$, as function of the charging energy, 
$U$, obtained using Eq.~(\ref{gamma_u_t_asym}) with $\epsilon_d=-U/2$, and NRG solutions 
for the exact system- and energy-dependent hybridization function. The corresponding 
electrodes-molecule configurations are illustrated in Fig.~\ref{fig:fig5}. 
The horizontal orange region illustrates the experimental range of Kondo 
temperatures of about 1-3 K~\cite{frisenda2015kondo}.
}
	\label{fig:TK_u}
\end{figure}
In order to obtain a first estimate of the Kondo temperature for different junctions presented in Fig.~\ref{fig:fig5}, we 
assume a constant (energy independent) coupling $\Gamma =\Gamma(E_\mathrm{F}=0)$. The Anderson model then maps onto the Kondo model while approaching the so-called local moment limit, where $|\epsilon_d| \gg \Gamma $ and $|\epsilon_d +U| \gg \Gamma $, $(n_d \approx 1)$~\cite{schrieffer1966relation} (see ESI Sec. S3 for details), and the Kondo temperature is given by the Haldane equation~\cite{Hewson.book,Haldane}
\begin{equation} 
k_\mathrm{B} \theta_\mathrm{L}=\frac{1}{2}\sqrt{\Gamma U} e^{-\frac{\pi \left|\epsilon_d\right|  \left|\epsilon_d+U \right|}{U \Gamma}},\label{gamma_u_t_asym}
\end{equation} 
with $-U\le\epsilon_d\le0$.
The results obtained with this expression are shown in Fig. \ref{fig:TK_u}, and are compared with the NRG calculations in the next subsection.
The experiments in Ref.~\citenum{frisenda2015kondo} show that the SOMO of the PTM is close to half-filling and that it can be brought to exact half-filling by applying a gate voltage to the system. Here we therefore consider only this half-filled case, and the effects of small deviations from half filling are presented in Sec. \ref{sec:super-p}. We note that if the molecule is partially charged, then in general $\theta_\mathrm{L}$ increases compared to the charge neutral state \cite{Droghetti.17}, so that the values for half filling represent a lower limit for the theoretical results.
The values of $\Gamma$ for each structure are taken from Fig. \ref{fig:fig5}, and at half-filling for a particle-hole symmetric SIAM we have $\epsilon_d=-U/2$. 
For large enough $U$ all curves in Fig. \ref{fig:TK_u} decay exponentially with $U$, and the slope of the exponential decay is inversely proportional to $\Gamma$, so that the configurations with the largest $\Gamma$ have the slowest decay, and therefore the highest $\theta_\mathrm{L}$, for a given value of $U$. 

Experimentally it is found that $\theta_\mathrm{L}$ is approximately constant upon stretching of the junction, which indicates a highly asymmetric
coupling, where the molecule preserves the contact geometry to one electrode, while the contact with the other electrode is elongated. In other words, the molecule is strongly bound to one of the electrodes, which may correspond to the core of the PTM-BT lying flat on a rough Au surface with the thiophene linkers bridging both sides of the junction.
This conclusion is supported by calculations for the CFGs B2, B4 and T4 structures, which have the largest values of $\Gamma$, and which all show asymmetric couplings. The calculated $\theta_\mathrm{L}$ values lie in the experimental range if $U$ is equal to about 1~eV. This charging energy is considerably smaller than the gas phase value of about $4$ eV, and we ascribe this reduction of $U$ to the charge screening by the electrons in the Au surface (see ESI Sec. S2).
A value for the change of $U$ due to screening can be calculated using a number of methods \citep{Himmetoglu}, for molecules on general corrugated and irregular metal surfaces constrained DFT (cDFT) has been shown to give good results \cite{Amaury,Amaury2}. Alternatively, here we estimate it by approximating the metal surface as a plane, and by using a classical image charge model with a molecule between two metal electrodes ~\cite{Amaury} to capture this effect. In this way we calculate that a gap reduction of about $3$ eV corresponds to ideal planar Au electrodes at a distance of about 2.7 \AA~from the center of the molecule. This number is similar to the distance of 3.4 \AA~for the CFG B2 structure. The remaining difference can be due to either an overestimated theoretical gas phase gap, or due to the experimental atomic structures having an even stronger binding between molecule and electrodes than CFG B2.

For the structures with small $\Gamma$ the value of $U$ that brings $\theta_\mathrm{L}$ in the experimental range is very small, and goes below the expected possible range. Such junctions are therefore expected to exhibit a $\theta_\mathrm{L}$ well below the experimentally accessible temperatures. This is consistent with the experimental evidence that only a fraction of the molecular junctions, which we attribute to those with the largest $\Gamma$, exhibit a Kondo state at an experimentally accessible temperature. Overall our results confirm that the molecule lies flat on a rough Au surface when it exhibits Kondo behavior, since only such structures allow for small binding distances and strong electronic coupling.

\subsection{NRG calculations}
\label{subsec:NRG}
\begin{figure}
	\centering
   \includegraphics[width=0.32\textwidth,clip=true]{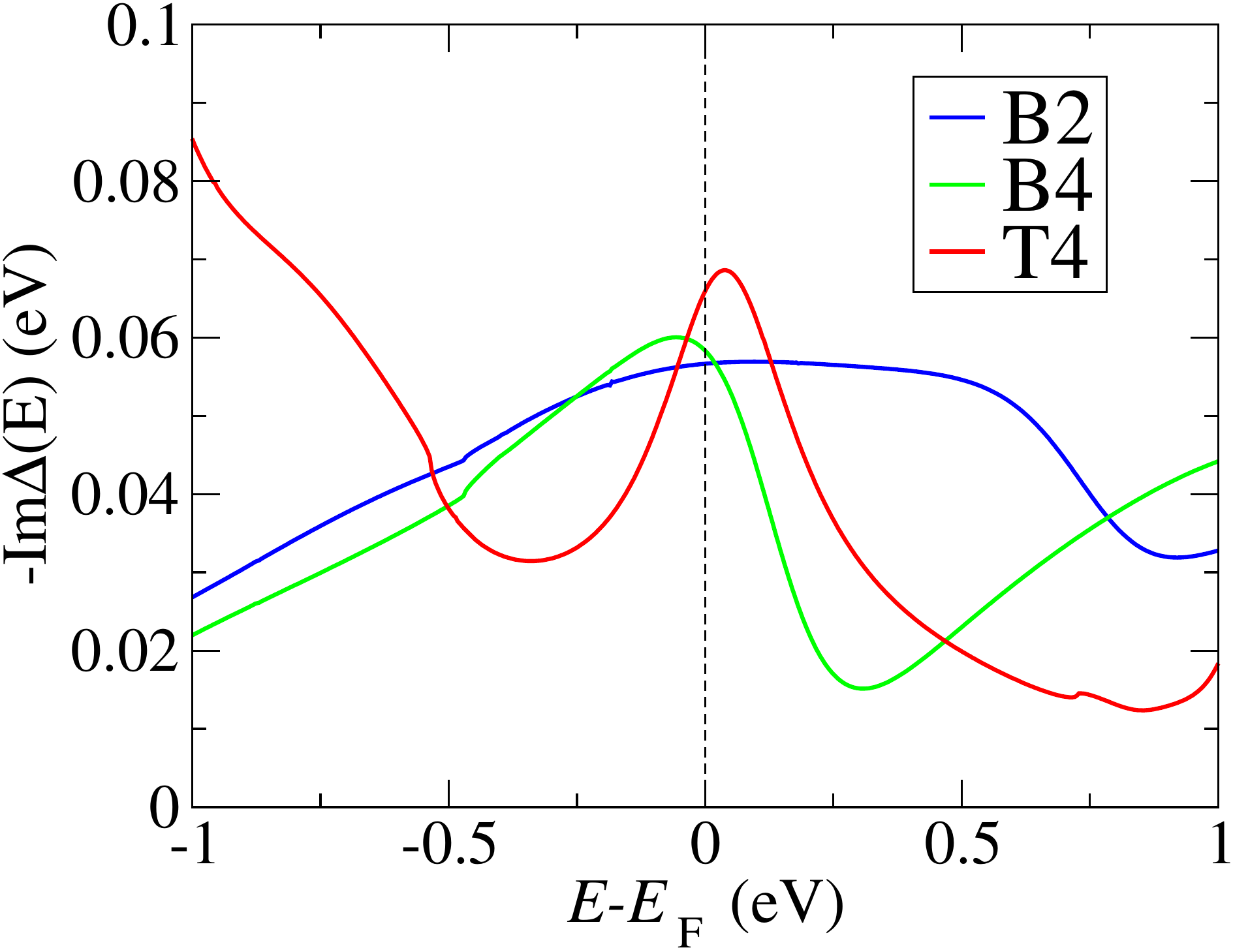}
\caption{Negative imaginary part of the hybridization function calculated using the DFT+NEGF method, and used as input for the NRG calculation, for the three configurations with the largest hybridizations (B2, B4, T4) (see Fig. \ref{fig:fig5}). High energy contributions are truncated as  outlined in the ESI Sec. S4.}
	\label{fig:fig8}
\end{figure}

In order to confirm the trends for $\theta_\mathrm{L}$ obtained with the simplified model Eq. (\ref{gamma_u_t_asym}), and to evaluate the conductance in the presence of electronic correlations,
we integrate NRG calculations in the method. We consider the SIAM representing the PTM/Au structures with the largest $\Gamma$ (CFGs B2, B4 and T4), for which in the previous subsection have estimated the Kondo temperatures to lie in the experimental range.
For each junction we calculate $\textrm{Im}\Delta(E)$ in Eq.~(\ref{imdelta}) by using DFT+NEGF with the method presented in Ref. \citenum{Droghetti.17}, and the results are shown in Fig.~\ref{fig:fig8}. While the value around $E_\mathrm{F}$ is similar for all cases, there are pronounced differences in the energy dependence. The NRG calculations then allow us to verify whether the approximation of a constant $\Gamma$ used so far is applicable for these realistic atomic structures. The real part is obtained with the Kramers-Kronig relation, Eq.~(\ref{redelta}). Further details about the NRG calculations are presented in the ESI Sec. S4.
The many-body self-energy calculated with NRG is then used to evaluate the zero-bias and zero-temperature transmission in the presence of strong correlations in the next section.

\begin{figure}
	\centering
         \includegraphics[width=0.45\textwidth,clip=true]{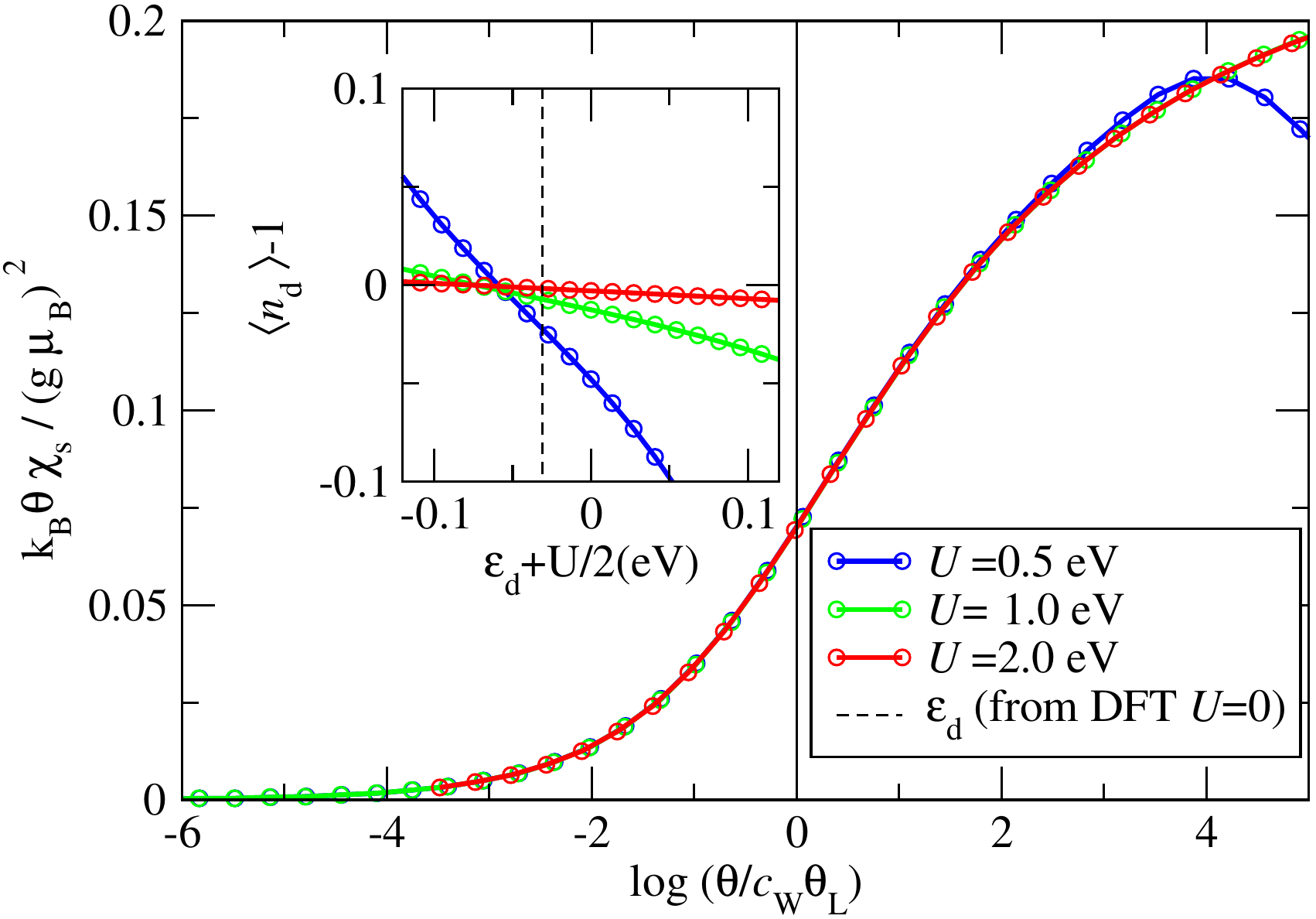} 
\caption{%
Main graph: The universal function displaying the temperature dependent scaling function  $(g \mu_B)^2F(\theta/c_\mathrm{W}\theta_\mathrm{L} )=k_\mathrm{B} \theta \chi_s$ 
against $\log(\theta/c_\mathrm{W}\theta_\mathrm{L})$. Inset: The impurity occupation as a function of the rescaled on-site energy $\epsilon_d+U/2$.  The on-site energy obtained within DFT is indicated as the vertical dashed line.
}
	\label{fig:fig9}
\end{figure}

The Kondo temperature is extracted from the impurity contribution to the magnetic susceptibility $\chi_\text{s}(\theta)$ \cite{PhysRevB.21.1003} (see ESI Sec. S5).
 In Fig.~\ref{fig:fig9} we present $\chi_\text{s}(\theta)$ for the B4 geometry, where the inset shows the small deviation of the impurity occupation $n_d$ from the half-filled case ($n_d=1$). 
We find that $\chi_s(\theta)$ always follows the same universal behavior as long as the interaction strength is large enough ($U>0.5\,$eV). A crossover is observed from the high $\theta$ local moment regime, where $k_\mathrm{B}\theta \chi_s/(g\mu_\mathrm{B})^2=1/4$,
to the low $\theta$ strong correlation limit, where $k_\mathrm{B}\theta \chi_s/(g\mu_\mathrm{B})^2=0$; here $g$ is the Land\`e-factor and $\mu_\mathrm{B}$ the Bohr magneton~\cite{PhysRevB.21.1003}.
We find that for $U$ values above  $U=0.5\,$eV the curves can be collapsed onto a single universal function. Note that for $U=0.5\,$eV one can already recognize the deviation from the universal behavior as a dip in the high $\theta$ susceptibility.

The collapse of the susceptibilities is interpreted as a universality due to the formation of a Kondo-singlet.
In the local moment regime the static spin-susceptibility scaled by the Kondo-temperature follows the same universal curve~\cite{PhysRevB.21.1044}, where the scaling function $F(x)$ is defined by~\cite{Hewson.book}
\begin{equation}
\frac{k_\mathrm{B}\theta\chi_s}{(g\mu_\mathrm{B})^2}=
F\left(\frac{\theta}{c_\mathrm{W}\theta_\mathrm{L}}\right),
\label{eq:scale}
\end{equation}
and where $c_\mathrm{W}$ is the so called Wilson number, which is a model-dependent constant (see ESI Sec. S3).
Here, the Kondo temperature $\theta_\mathrm{L}$ plays the role of a scale invariant in the renormalization group (RG) language. This means that systems with different initial parameters end up in the same low temperature fixed point after mode elimination (RG-flow towards the same fixed point)~\cite{Hewson.book}.
This gives rise to  the universal behavior in Fig.~\ref{fig:fig9} at low $\theta$.
The value for $\theta_\mathrm{L}$ is obtained in the standard way from the condition that the universal function at $\theta=c_\mathrm{W}\theta_\mathrm{L}$ is $ F(1)=0.07$~\cite{PhysRev.124.41}. 
In Fig.~\ref{fig:TK_u} the values of $\theta_\mathrm{L}$ calculated in this way are displayed as dashed lines. Importantly, they agree rather well with those obtained using the approximate Eq.~(\ref{gamma_u_t_asym}), showing that the approximation of a constant $\Gamma$ is valid for this system.
The NRG results therefore also confirm the conclusion that for the three structures with large $\Gamma$ the value of $\theta_\mathrm{L}$ is in the experimental range for $U\approx 1$ eV.

\section{Electron transmission}
\label{subsec:trans}

To evaluate the transport properties of this system we add the zero-temperature NRG self-energy, $\Sigma(E,\theta=0)$, to the DFT+NEGF Green's function via the Dyson equation and compute the resulting energy-dependent transmission function, $T_\mathrm{t}(E, \theta=0)$, in the presence of many-body correlations not captured at the standard DFT-KS level~\cite{Jacob5,Liviu,Droghetti.17}. As outlined in the ESI Secs. S7
 and
S8,
the linear response zero-temperature conductance, $G_0=G(V=0,\theta=0)=dI(V,\theta=0)/dV|_{V=0}$, is given by 
\begin{equation}
G_0=\frac{2 e^2}{h}T_\mathrm{t}(E_\mathrm{F},\theta=0),\label{EqG0}
\end{equation}
where $e$ is the electron charge, $h$ the Planck constant and $2e^2/h$ the quantum of conductance. Note that we have also implicitly assumed that there is no external magnetic field, whose effect will be considered in the next section. When $\Gamma_\mathrm{L}\gg\Gamma_\mathrm{R}$ ($\Gamma_\mathrm{L}\ll\Gamma_\mathrm{R}$) this can be extended to finite $V$ as $G(V,0)\approx(2 e^2/h)T_\mathrm{t}(-e V,0)$ ($G(V,0)\approx(2 e^2/h)T_\mathrm{t}(+e V,0)$). As discussed in the previous sections, we expect the Au/PTM/Au system exhibiting Kondo behavior to have such a highly asymmetric coupling. This condition is indeed fulfilled for CFGs B2 and B4, and to a minor extent also for CFG T4, so that the energy dependence of the transmission approximately corresponds to the voltage dependence of the conductance.
Note that for such highly asymmetric coupling the dominant effect of the voltage is a shift of the molecular energies due to its induced local electric field, while for the case of approximately symmetric coupling ($\Gamma_\mathrm{L}\approx\Gamma_\mathrm{R}$) the current induced non-equilibrium change of occupation gives an additional important contribution and therefore needs to be taken into account.
In the next section we will therefore generalize these relations and provide the non-equilibrium relations for the conductance that are also valid for arbitrary values of $\Gamma_\mathrm{L}$ and $\Gamma_\mathrm{R}$.

As outlined in Ref. \citenum{Droghetti.17} and in the ESI Sec.~S7,
the total transmission function is the sum of the elastic transmission, $T$, and of the inelastic impurity transmission, $T_\mathrm{R,AI}$, so that $T_\mathrm{t}=T+T_\mathrm{R,AI}$. The elastic transmission has contributions from electrons flowing through the impurity, $T_\mathrm{AI}$, from the background transmission, $T_\mathrm{B}$, and from interference terms, $T_\mathrm{I}$ ($T=T_\mathrm{AI}+T_\mathrm{B}+T_\mathrm{I}$).
Notably, at zero-temperature, for a system in the Kondo regime one has 
$T_\mathrm{R,AI}(E_\mathrm{F})=0$, because the imaginary part of the impurity many-body self-energy vanishes at $E_\mathrm{F}$ in accordance with the 
Fermi-liquid picture~\cite{Hewson.book}.
\begin{figure}
\includegraphics[width=0.48\textwidth,clip=true]{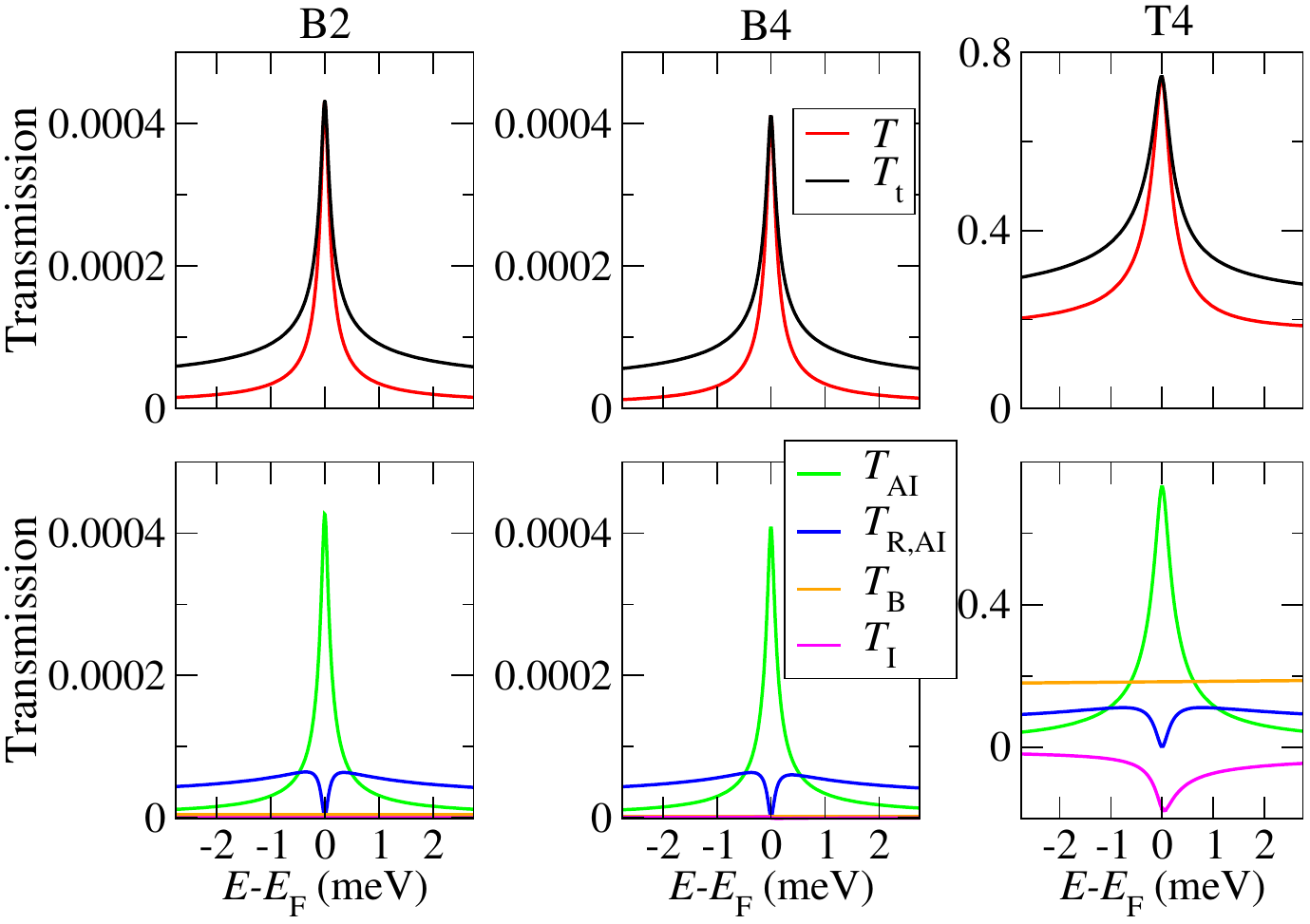}
\caption{Zero-bias transmission including the zero-temperature NRG self-energy for the B2, B4 and T4 structures, and for $U=1.0$ eV. Here $T$ is the total coherent transmission, 
$T_\mathrm{t}$ is the total transmission including incoherent effects, $T_\mathrm{AI}$ is the coherent transmission component of the AI itself, $T_\mathrm{B}$ is the coherent background transmission, $T_\mathrm{I}$ is the interference term, and $T_\mathrm{R,AI}$ is the incoherent transmission. The total transmission is then $T_\mathrm{t}=T+T_\mathrm{R,AI}$, with $T=T_\mathrm{AI}+T_\mathrm{B}+T_\mathrm{I}$. Note the different 
scales of the transmission-axis for B2, B4 and T4.} \label{fig:TRC_B2_B4_T4}
\end{figure}

The calculated low energy transmissions for the B2, B4 and T4 configurations are presented in Fig.~\ref{fig:TRC_B2_B4_T4}. 
Here $U$ is set to $1$ eV, since this is the charging energy that provides a Kondo temperature in the experimental range. 
The results for different values of $U$ are shown in the ESI Sec. S7.
One can clearly identify the Kondo peak around $E_\mathrm{F}$, which has a width of the order of 1 meV, in good agreement with the experiments \cite{frisenda2015kondo}.
The overall dominant contribution comes from $T_\mathrm{AI}$ for all cases.
While the background transmission and interference terms are negligible in the highly asymmetric setups (CFGs B2 and B4), they are rather large in the break-junction geometry T4. 
Importantly, while in the B2 and B4 geometry the transmission values are very small, for the T4 break-junction configuration they can reach values up to $0.8$, and such variations are indeed found in experiments \cite{frisenda2015kondo}. 
In the present case the magnitude of both the background transmission and of
the Kondo peak become large for symmetric coupling ($\Gamma_\mathrm{L}\approx\Gamma_\mathrm{R}$), while they progressively decrease as the coupling becomes more asymmetric. However,
we point out that the background transmission may generally be very large if the overlap between the Au electrodes is large or if the electrodes are very broad. 
In that case one may still have $\Gamma_\mathrm{L}\gg\Gamma_\mathrm{R}$ for the molecule itself, 
but the background current will be much larger than that flowing through 
the molecule. 
Therefore, for a comparison between theory and experiments for the Kondo conductance itself
ideally one needs to separate out the background conductance. While this is difficult to do in experiment, our simulation scheme allows to perform this separation for each atomic configuration.
In Fig.~\ref{fig:TRC_B2_B4_T4} we also plot the incoherent 
transmission $T_\mathrm{R,AI}$ and $T_\mathrm{t}=T+T_\mathrm{R,AI}$, which determines the measured conductance.
As stated above, $T_\mathrm{R,AI}$ vanishes at $E_\mathrm{F}$, while it leads to a further overall enhancement of the transmission spectrum away from it. It therefore does not affect the zero-bias and zero-temperature conductance, but it plays an important role at finite bias and finite temperatures, as discussed below.

Although the results shown so far are obtained for zero temperature, we can obtain an estimate of the temperature dependence of the full width at half maximum (FWHM, $W$) of the Kondo peak in the DOS by performing a low energy expansion of the SIAM DOS. For the system investigated here we consider the half-filled particle-hole symmetric case, and moreover, since $\Gamma \ll U$, we are in the so-called strong correlation regime \cite{Hewson.93}. As shown in the ESI Sec. S6,
in such a regime the dependence of the FWHM on temperature for a SIAM with energy-independent hybridization $\Delta=\Gamma/2$ is approximately given by
\begin{equation}
W(\theta,\tilde{\Delta})=
\tilde\Delta 2 \sqrt{2}\sqrt{\sqrt{1+\left(\frac{\pi^2 k_\mathrm{B}^2 \theta^2}{2 \tilde\Delta^2}+1\right)^2}-1}.
\label{eq:fwhm}
\end{equation}
Here $\tilde{\Delta}$ is the renormalized quasi-particle spectral width, $\tilde{\Delta}=z \Delta$, and $z=\left[1-\partial_E\,\Re{\left(\Sigma(E,\theta=0)\right)}_{E=E_\mathrm{F}}\right]^{-1}$ is the so called wave-function renormalization factor~\cite{Hewson.93,Munoz.11}. Note that here we use the zero temperature limit of $\Sigma(E,\theta)$, since we perform the perturbation expansion around $\theta=0$, but in general $z$ can also be evaluated at finite temperature by using the finite-temperature $\Sigma_\sigma(E,\theta)$ in its definition above.
Furthermore, in the particle-hole symmetric regime $\tilde{\Delta}$ is related to the Kondo temperature as~\cite{Hewson.93}
\begin{equation}
k_\mathrm{B}\theta_\mathrm{L}=\frac{\pi}{4}\tilde{\Delta}.
\label{eq:thetal}
\end{equation}
Note that the relation in Eq. (\ref{eq:fwhm}) is different from the widely used form given in Ref. \citenum{Nagaoka.02}, since in that reference the energy dependence of the real part of the many-body self-energy is neglected. In the ESI Sec. S6
 we show that $W(\theta,\tilde{\Delta})$ from Eq. (\ref{eq:fwhm}) reproduces rather well the NRG results up to temperatures of about $2 \theta_\mathrm{L}$.

In experiments, $\theta_\mathrm{L}$ can be obtained by fitting Eqs. (\ref{eq:fwhm}) and (\ref{eq:thetal}) to the measured temperature dependent data for the FWHM of the conductance peak. Note that this is somewhat larger than the FWHM of the DOS due to the additional temperature induced broadening of the Fermi distribution of the electrons (see ESI Sec. S8).

\begin{figure}
\includegraphics[width=0.4\textwidth,clip=true]{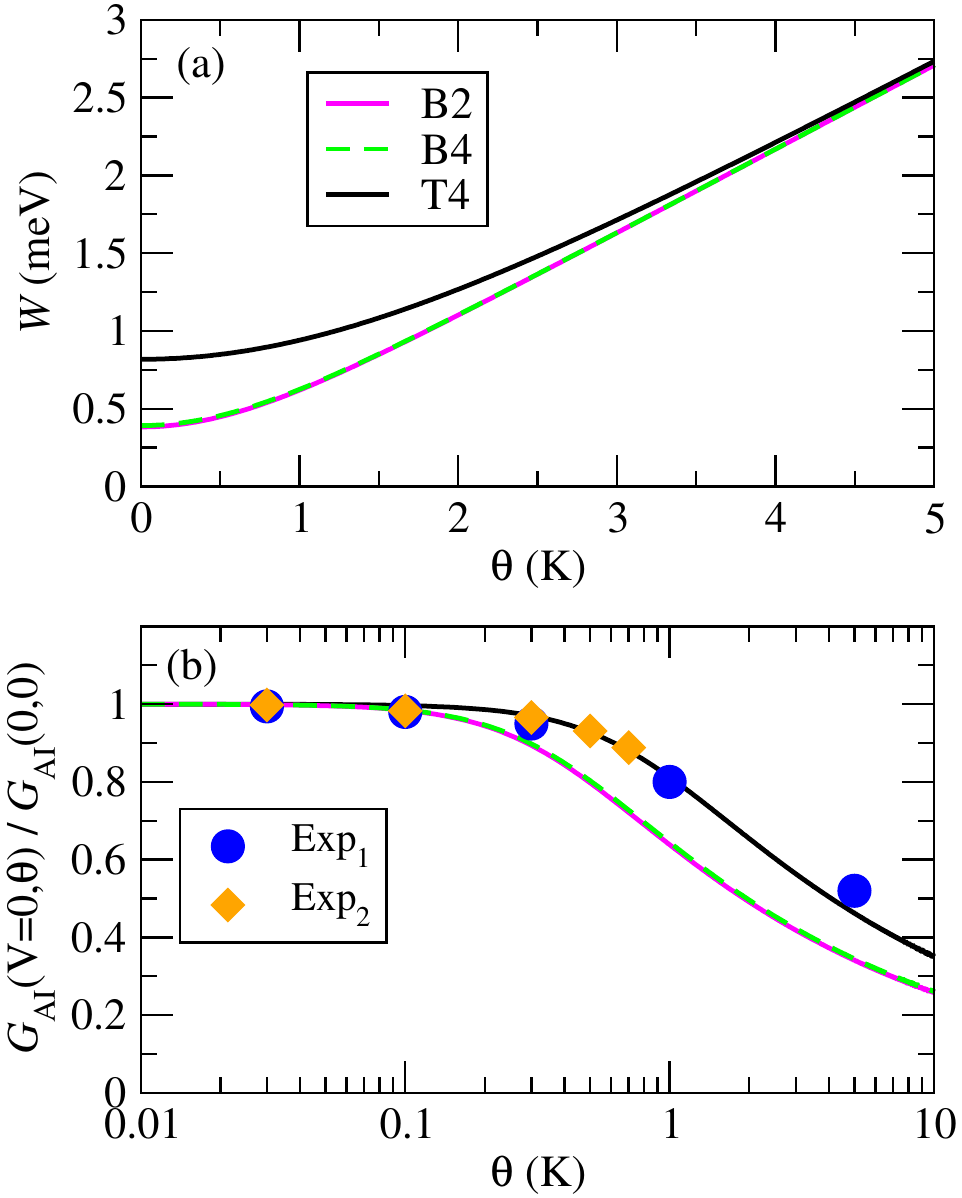}
\caption{(a) Full width at half maximum of the Kondo peak in the DOS, $W$, calculated with the model Eq. (\ref{eq:fwhm}); (b) normalized impurity conductance as function of temperature using the function in Eq. (\ref{eq:GHG}), for the B2, B4 and T4 configurations. The results are compared to experimental data from Ref. \citenum{frisenda2015kondo}, denoted as ``Exp$_1$'' and ``Exp$_2$''.} \label{fig:WofT}
\end{figure}

From our NRG calculations we can extract effective values of $\tilde\Delta$ for 
the three configurations B2, B4 and T4, which take into account the energy-dependent hybridization at an average level (see ESI Sec. S6).
The resulting values, together with the corresponding Kondo temperatures, are shown
in Table \ref{tab:deltatilde}. Note that the values for $\theta_\mathrm{L}$ calculated in this way are in good agreement with the values calculated directly from the NRG susceptibility (Fig. \ref{fig:TK_u}). In Fig. \ref{fig:WofT}(a) we present the resulting temperature dependent FWHM for all three systems calculated using Eq. (\ref{eq:fwhm}) and the parameters in Table \ref{tab:deltatilde}.
\begin{table}[t]
\centering
\begin{tabular}{ c | c | c| c }
& B2 & B4 & T4 \\
\hline
$\theta_\mathrm{L}$ (K)       & 1.91    & 1.96    & 4.09  \\
$\tilde\Delta$ (meV) & 0.210   & 0.215   & 0.449 \\
$z$                  & 0.00315 & 0.00326 & 0.00598 \\
$U/\pi\Delta$   & 5.63 & 5.44 & 5.05 \\
\hline
\end{tabular}
\caption{Renormalized quasi-particle spectral width, $\tilde\Delta$, and corresponding Kondo temperature $\theta_\mathrm{L}$ calculated with Eq. (\ref{eq:thetal}), as well as wave-function renormalization factor, $z$, for the configurations B2, B4, T4 (note that $\tilde\Delta$ and $z$  given here are denoted as $\tilde\Delta_{\mathrm{\Sigma}}$ and $z_{\mathrm{\Sigma}}$ in the ESI Sec. S6 and in Table S1).
For $U=1$ eV we also give the value of $U/\pi\Delta$, with the values of $\Delta=\Gamma/2$ taken from Fig. \ref{fig:fig5}.}\label{tab:deltatilde}
\end{table}

Finally, within the approximation considered in this section we also estimate the temperature dependence of the normalized conductance of the Anderson impurity at zero-bias.
If one neglects the interference terms ($T_I\approx 0$), then one can write $G(V,\theta)\approx G_\mathrm{AI}(V,\theta)+G_\mathrm{B}(V,\theta)$, where $G_\mathrm{B}$ is the background conductance originating from $T_\mathrm{B}$, and $G_\mathrm{AI}$ is the conductance due to $T_\mathrm{AI}+T_\mathrm{R,AI}$. The temperature dependence of $G_\mathrm{B}$ is usually small, and for small $V$ also the voltage dependence can be neglected, so that we set $G_\mathrm{B}$ to be a constant background conductance. Within the approximations used in this section, the temperature dependence of $G_\mathrm{AI}$ is derived in the ESI Sec. S8 (Eq.~(S29) of the ESI) to be
\begin{eqnarray}
\frac{G_\mathrm{AI}(0,\theta)}{G_0}&\approx&1-\frac{\pi^4}{16}\left(\frac{\theta}{\theta_\mathrm{L}}\right)^2=1-\pi^2\left(\frac{k_\mathrm{B}\theta}{\tilde\Delta}\right)^2,
\label{eq:g0theta}
\end{eqnarray}
where $G_0=G_\mathrm{AI}(0,0)$.
If accurate experimental data are available at low $\theta$, then the mapping of the measured temperature dependent conductance profile to this equation allows to determine the experimental $\theta_\mathrm{L}$. However, in many experiments including also those for Au/PTM/Au junctions in Ref. \citenum{frisenda2015kondo}, the low temperature conductance data is too noisy, so that $\theta_\mathrm{L}$ is estimated from the high temperature data. Since no analytic expression is available for the whole temperature range, in Ref.~\citenum{Goldhaber-Gordon.98} a functional form is introduced in order to fit calculated NRG results in Ref. \citenum{Costi.94c}. The proposed fitting curve is
\begin{equation}
\frac{G_\mathrm{AI}(0,\theta)}{G_0}=\left(\dfrac{1}{1+\big( \theta/\tilde{\theta}_\mathrm{K} \big)^2}  \right)^{s},
\label{eq:GHG}
\end{equation}
where $\tilde{\theta}_\mathrm{K}=(2^{1/s}-1)^{-1/2}\theta_\mathrm{K}$, and $\theta_\mathrm{K}$ and $s$ are phenomenological 
parameters. The value of $\theta_\mathrm{K}$ sets the temperature at which the conductance is reduced by a factor 2 ($G_\mathrm{AI}(0,\theta_\mathrm{K})=G_\mathrm{AI}(0,0)/2$). The second order expansion of this relation leads to $G_\mathrm{AI}(0,\theta)/G_\mathrm{AI}(0,0)\approx 1-s(2^{1/s}-1)\left( \frac{\theta}{\theta_\mathrm{K}}\right)^2$. As outlined in the ESI Sec. S8,
for the particle-hole symmetric SIAM one can approximate $\theta_\mathrm{K}\approx\theta_\mathrm{L}$. Furthermore, the condition that the second order expansion needs to be equal to the form given in Eq. (\ref{eq:g0theta}) then sets the value of $s$ to be $s\approx0.20$. 

A comparison of the temperature dependent conductance obtained using Eq. \ref{eq:GHG} for the B2, B4 and T4 structures with the experimental data in Ref. \citenum{frisenda2015kondo} is plotted in Fig.~\ref{fig:WofT}(b).
The experimental normalized conductance
agrees rather well with the calculated curves, in particular with the one for T4, which has the highest Kondo temperature of all the calculated structures.
We denote as ``Exp$_1$'' and ``Exp$_2$'' the data for the two sets of experiments presented in Figs.~3c and 3d of Ref.~\citenum{frisenda2015kondo}, respectively. 
When extracting the experimental Anderson Impurity conductance one has to first subtract the background conductance, $G_\mathrm{B}$.
Our calculations show that the background conductance depends significantly on the detailed atomic structure, as shown by the values of $T_\mathrm{B}$ in Fig.~\ref{fig:TRC_B2_B4_T4}. 
However, in experiments only the total 
conductance is accessible. One can approximate the background conductance by the conductance at zero bias for a very large applied magnetic field, which can be extracted from Figs. 3g-h of Ref. \citenum{frisenda2015kondo}. In this way we extract the ratio of background conductance to the total conductance at zero bias and zero temperature to be about 0.29 for Exp$_1$, and 0.34 for Exp$_2$.

While the results presented in this section show good agreement with the experiments in Ref. \citenum{frisenda2015kondo}, the limitation is that the equations are all based on the assumption of a particle-hole symmetric system, which is not generally the case. Indeed, in Ref. \citenum{frisenda2015kondo} it is also shown that by applying a gate voltage the occupation of the SIAM can be systematically changed.
At particle-hole symmetry the system is characterized by a single energy scale, $k_\mathrm{B}\theta_\mathrm{L}$, and Eqs. (\ref{eq:fwhm}-\ref{eq:GHG}) reflect this property.
Away from particle-hole symmetry, however, this  no longer holds and corrections 
to these formulas enter.
Furthermore, the condition that $\Gamma_\mathrm{L}$ is very different from $\Gamma_\mathrm{R}$ does not apply for a general system.
In the next section we will therefore extend the method to the general non-equilibrium case, and also to the case away from particle-hole symmetry within a perturbative approach. 

\section{Non-equilibrium relations: renormalized super-perturbation theory}
\label{sec:super-p}

In this section we account for finite-temperature ($\theta > 0$) and general finite-bias ($V\neq 0$) effects by using the renormalized super perturbation theory (rSPT) described in Refs.~\citenum{Munoz.11,Merker.13,Munoz.16}. The rSPT corresponds to a perturbative method organized around the particle-hole symmetric strong coupling fixed point considered in the previous sections. While for the PTM/Au system considered here we always have $U\gg\Delta$, the rSPT relations are in principle valid for arbitrary values of $U$, and  account for deviations from the particle-hole symmetry at a perturbative level.  
It is based on the insight that at the strong-coupling fixed point the equations have the form of an Anderson model, albeit with renormalized parameters \cite{Hewson.93}. These parameters are the renormalized hybridization, $\tilde\Delta$, which has been introduced in the previous section ($\tilde\Delta=z \Delta$), the renormalized energy level, $\tilde\epsilon_d$, which is given by $\tilde{\epsilon}_d =(\epsilon_d + U/2)/\Delta$, and the renormalized interaction energy, $\tilde{U}$, defined in the ESI Sec. S9.
We introduce the rescaled renormalized interaction $\tilde{u}=\tilde{U}/{\pi\tilde\Delta}$, which lies in the range from 0 for small $U$ to 1 for very large $U$ (see Fig. S9).
In this section we present results as function of $\tilde{\epsilon}_d$, which determines the deviation from the particle-hole symmetric case, and which can be tuned experimentally by applying a gate voltage \cite{frisenda2015kondo}.

The Kondo temperature $\theta_L$ near the strong coupling fixed point is obtained as
$k_\mathrm{B} \theta_\mathrm{L}=\left((g\mu_\mathrm{B})^2/4\right)\lim_{\theta \rightarrow 0} \big(\chi_\mathrm{s}\big)^{-1}$ \cite{Merker.13},
with the $\theta=0$ limit of the static spin susceptibility\cite{Hewson.93}
\begin{eqnarray}
\lim_{\theta \rightarrow 0} \chi_\mathrm{s} = \frac{(g\mu_\mathrm{B})^2}{2}\tilde{A}_\mathrm{AI}(0,0)\left(1 + \tilde{U}\tilde{A}_\mathrm{AI}(0,0) \right),
\label{eq:chirspt}
\end{eqnarray}
and where $\tilde{A}_\mathrm{AI}(E=0,\theta=0)=z^{-1} A_\mathrm{AI}(E=0,\theta=0)$ denotes the equilibrium quasi-particle renormalized spectral density at the Fermi energy. Note that for the particle-hole symmetric reference system this definition of $\theta_\mathrm{L}$ is equivalent to the one presented in Sec. \ref{subsec:NRG} (see also ESI Sec. S3).
Up to second order in $\tilde{u}\tilde\epsilon_d$ we have
\begin{equation}
\tilde{A}_\mathrm{AI}(0,0)\approx \left[\pi\tilde\Delta\left(1+(1-\tilde{u})^2\tilde\epsilon_d^2\right)\right]^{-1}.
\end{equation}
Inserting this into Eq. (\ref{eq:chirspt}) yields the Kondo temperature 
\begin{eqnarray}
k_\mathrm{B} \theta_\mathrm{L}&=&\frac{2+2(1-\tilde{u})^2\tilde{\epsilon}_d^2}{1 + \frac{\tilde{u}}{1+(1-\tilde{u})^2\tilde{\epsilon}_d^2}}  \frac{\pi\tilde{\Delta}}{4},
\end{eqnarray}
which is a
generalization 
to finite $\tilde\epsilon_d$ and to arbitrary $U$ of the result for the symmetric SIAM in the strong coupling limit given in Eq. (\ref{eq:thetal}).

\begin{figure}
	\centering
\includegraphics[width=0.45\textwidth,clip=true]{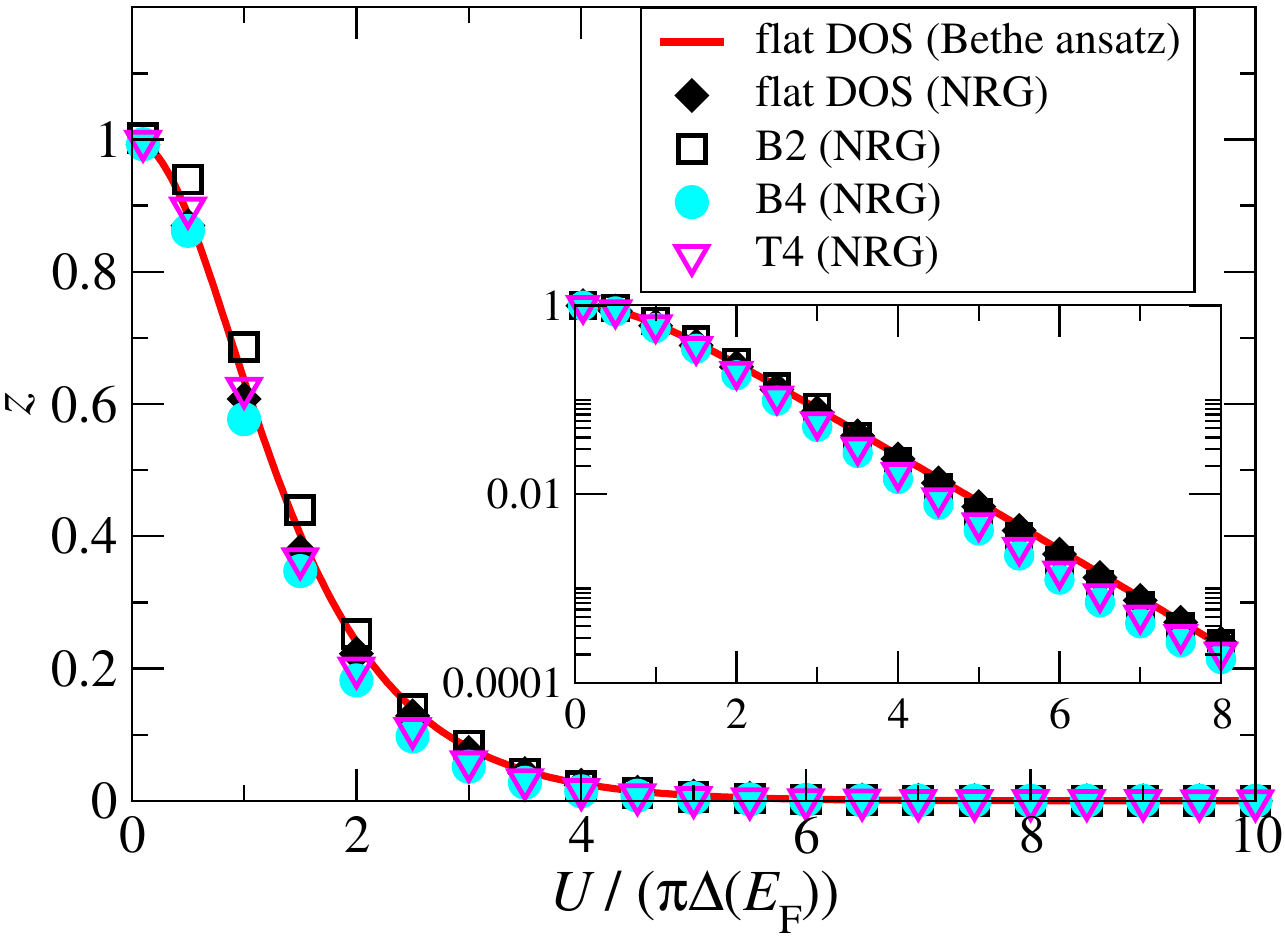}
\caption{
(Main graph) Comparison between the wave-function renormalization factor, $z$, calculated using NRG, for a constant hybridization function (black diamonds), and for the three configurations B2 (turquoise filled disks), B4 (pink open triangles), and T4 (white open rectangles). The Bethe ansatz solution for the same constant hybridization used in the NRG calculation is shown as the red solid line. Inset: Wave-function renormalization factor from the main graph plotted on a semi-logarithmic scale.
}
	\label{fig:fig10}
\end{figure}
A central issue is the relation between renormalized and bare parameters, which is  encoded in the wave-function renormalization factor $z$.
The renormalization factor $z=\tilde{\Delta}/\Delta$ can be obtained from NRG for a general energy-dependent hybridization function, and from Bethe ansatz for the case of a constant energy-independent hybridization function \cite{Hewson.93}.
A comparison between $z$ calculated for a constant hybridization function $\Delta(E)=\Delta(E_\mathrm{F})=\Gamma/2$ using Bethe ansatz and the NRG is shown as function of the interaction energy in Fig.~\ref{fig:fig10}, and demonstrates that they agree well. To address the question of the effect of an energy-dependent hybridization function on $z$, we also calculate $z$ using NRG and with the full energy-dependent $\Delta(E)$ for the B2, B4, and T4 structures (Fig \ref{fig:fig8}). Importantly, we find that they also agree rather well with the results for constant hybridization, showing that the low energy SIAM is largely dominated by the hybridization function around the Fermi energy.
Based on these results we therefore calculate $z$ and $\tilde{u}$ for the rSPT expansion using the Bethe ansatz for the particle-hole symmetric SIAM with constant hybridization $\Delta(E_\mathrm{F})=\Gamma/2$ (see Fig. S9 in the ESI Sec. S9), where the values of $\Gamma$ for each configuration are given in Fig. \ref{fig:fig5}.

In order to generalize the relations for the conductance, we first evaluate the equilibrium conductance, $G_0=G_\mathrm{AI}(V=0,\theta=0,B=0)$, defined in Eqs.~(\ref{EqG0}) and (S18) of the ESI, away from the particle-hole symmetry. Here we have explicitly noted that we consider the reference case with zero magnetic field ($B$). 
This results to
\begin{eqnarray}
G_0&=\dfrac{2e^2}{h} \dfrac{4\Gamma_\mathrm{L}\Gamma_\mathrm{R}}{\Gamma_\mathrm{L}+\Gamma_\mathrm{R}}2\pi A_\mathrm{AI}(0,0).
\end{eqnarray} 
Then the extension of rSPT to current-carrying steady states allows us to evaluate the non-linear low-voltage conductance for finite temperatures and also magnetic fields, which has the form \cite{Munoz.11,Merker.13,Munoz.16}
\begin{eqnarray} 
\frac{G_0 - G_\mathrm{AI}(V,\theta,B)}{G_0} &=& c_{\theta}\left(\frac{k_\mathrm{B}\theta}{\tilde\Delta} \right)^2 +c_B \left(\frac{|g|\mu_\mathrm{B} B}{\tilde\Delta} \right)^2 \nonumber\\ 
&& \hspace*{-80pt}  + c_{V}\left(\frac{eV}{\tilde\Delta} \right)^{2} 
-c_{V E_{d}}\left( \frac{eV}{\tilde\Delta} \right) \nonumber \\
&& \hspace*{-80pt} - c_{\theta  V} \left(\frac{eV}{\tilde\Delta}  \right)^{2}\left( \frac{\theta}{\tilde\Delta}  \right)^{2}
+c_{\theta VE_{d}} \left(\frac{eV}{\tilde\Delta}  \right)^{ }\left( \frac{\theta}{\tilde\Delta}  \right)^{2}. 
\label{eq:gTaylor}
\end{eqnarray} 
This result can be obtained by expanding $G_\mathrm{AI}(\theta,V,B)$ up to second order in $eV/\tilde{\Delta}$, $k_\mathrm{B} \theta/\tilde{\Delta}$, and $g \mu_\mathrm{B} B/\tilde{\Delta}$. The relations for the expansion coefficients are presented in the ESI Sec. S9, and extend the second order coefficients in $U$ given in Ref. \citenum{Munoz.11} to arbitrarily large values of $U$.
Note that the equilibrium transmission calculations, presented in the previous section and in the ESI Sec. S8, allow to extract the values of $c_{\theta}=\pi^2$ and also $c_V=3/2$ in the strong coupling limit ($\tilde{u}=1$) and at particle-hole symmetry ($\tilde\epsilon_d=0$), and for highly asymmetric coupling to the electrodes (Eq. (S29) in the ESI).
Using the general rSPT relations given in the ESI Sec. S9 one can see that as long as $\tilde{u}=1$ and $\tilde\epsilon_d=0$ these values are valid for arbitrary $\Gamma_\mathrm{L}$ and $\Gamma_\mathrm{R}$, so that they are independent of the level of asymmetry in the electronic coupling to the electrodes. Note that an important advantage of the rSPT approach is that it is not restricted to these limiting cases, and it is valid for arbitrary values of the parameters, which is a consequence of the fact that it is a truly non-equilibrium method.

\begin{figure}
	\centering
\includegraphics[width=0.47\textwidth,clip=true]{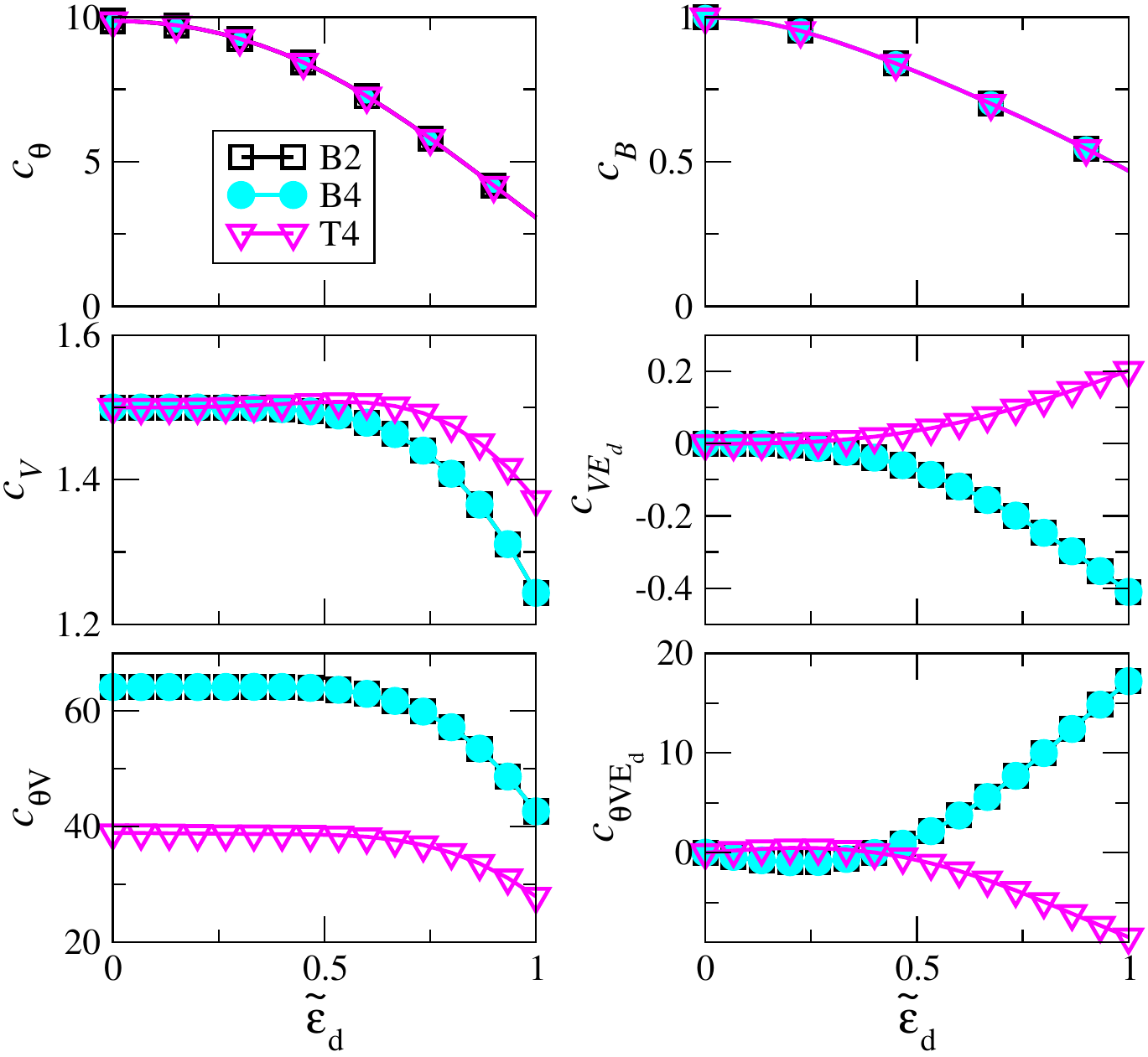}
\caption{The dependence of the conductance coefficients in Eq. (\ref{eq:gTaylor}) on the deviation from particle-hole symmetry, determined by $\tilde\epsilon_d=(\epsilon_d+U/2)/\Delta(E_\mathrm{F})$, for the configurations B2, B4 and T4. The mathematical relations for the coefficients are given in the ESI Sec. S9, and the parameter $\zeta = 3(\Gamma_\mathrm{L}/\Gamma_\mathrm{R})/(1 + \Gamma_\mathrm{L}/\Gamma_\mathrm{R})^2$ in those equations, which determines the asymmetry of the electronic coupling to the left and right electrodes, follows from the values of $\Gamma_\mathrm{R}$ and $\Gamma_\mathrm{L}$ in Fig. \ref{fig:fig5} as $\zeta_\mathrm{B2}=2.6\cdot 10^{-4}$, $\zeta_\mathrm{B4}=2.6\cdot 10^{-4}$, and $\zeta_\mathrm{T4}=0.568$. The
dimensionless Coulomb repulsion $U/(\pi\Delta(E_\mathrm{F}))$ for the effective Anderson model  applicable
to each system is presented in Table I.
}
	\label{fig:fig11}
\end{figure}

The rSPT expansion coefficients calculated for the B2, B4, and T4 structures are displayed in Fig.~\ref{fig:fig11} as a function of the local level energy $\tilde{\epsilon}_d = \left(\epsilon_d + U/2 \right)/\Delta$. As noted above, in an experiment this can be modified by applying a gate voltage. We use the Bethe ansatz $\tilde{u}$ for the values of $U/\pi\Delta$ give in Tab. \ref{tab:deltatilde}, which then result to $\tilde{u}=0.99999418$ for the B2 structure, $\tilde{u}=0.99999088$ for B4, and $\tilde{u}=0.99997705$ for T4. These values are all very close to 1, and indeed replacing them with 1 leads to essentially the same results, confirming that the Au-PTM system is in the strong coupling limit. The coefficients therefore differ only due to the changes in $\Gamma_\mathrm{L}/\Gamma_\mathrm{R}$, for which we use the DFT values given in Fig. \ref{fig:fig5}.
Since $c_{\theta}$ and $c_B$ are linear-response properties and do not depend on $\Gamma_\mathrm{L}/\Gamma_\mathrm{R}$, they are identical for all configurations. Consequently, $c_{\theta}$ and $c_B$ can also be calculated via NRG. A comparison for these two quantities between rSPT and NRG is given in Ref.~\citenum{Munoz.16}, where a rather good agreement is found up to moderate values of $\tilde\epsilon_d$.

\begin{figure}
	\centering
\includegraphics[width=0.46\textwidth,clip=true]{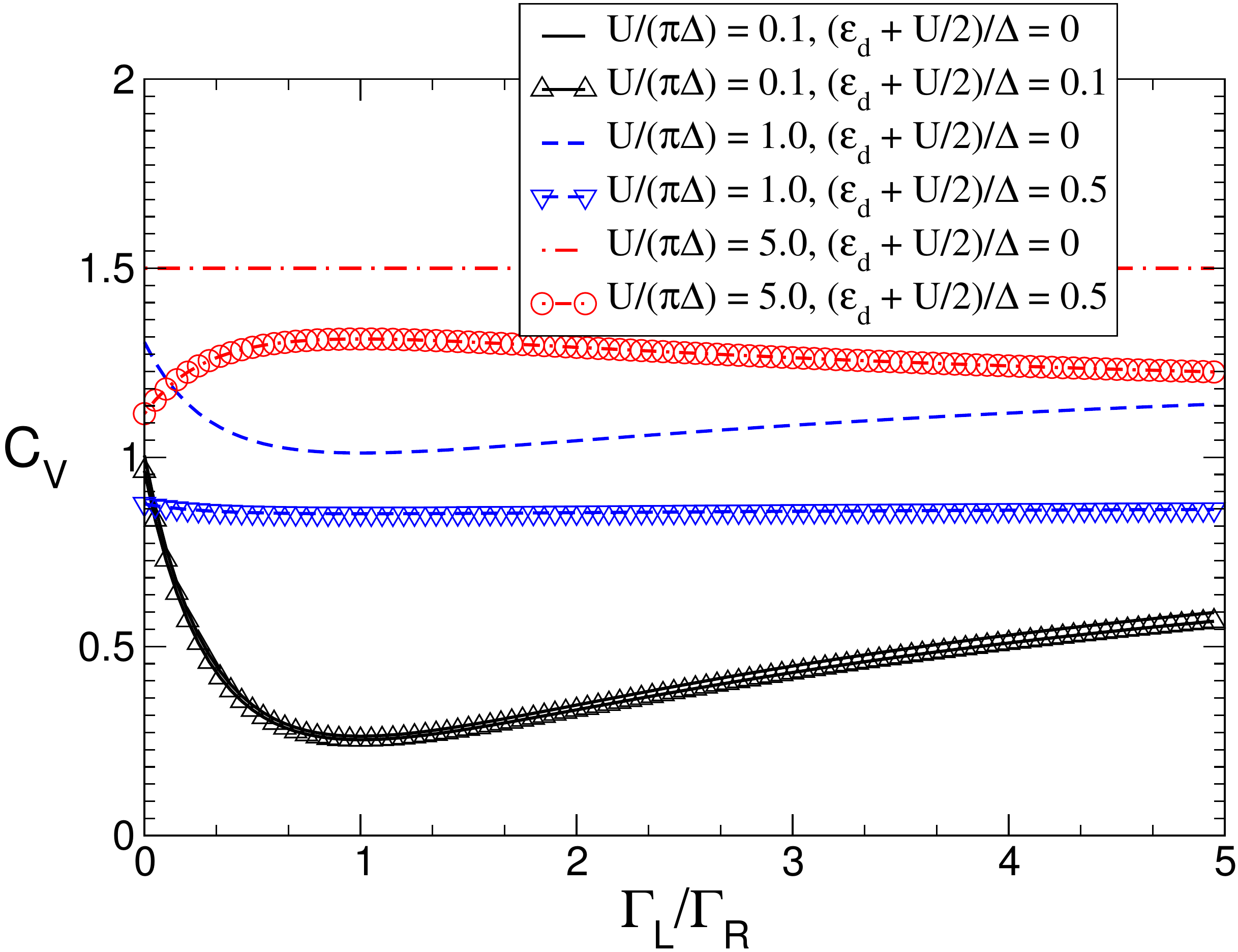}
\caption{The dependence of the transport coefficient $c_V$, obtained from the rSPT, is displayed as a function of the asymmetry in the contacts $\Gamma_\mathrm{L}/\Gamma_\mathrm{R}$, for different values of the dimensionless Coulomb repulsion $U/(\pi \Delta)$ and local
level energy $(\epsilon_d + U/2)/\Delta$.
}
	\label{fig:fig11b}
\end{figure}

The effect of the contact asymmetry, as captured by the ratio $\Gamma_\mathrm{L}/\Gamma_\mathrm{R}$, affects the value of the finite voltage coefficients, as clearly seen in the lower part of Fig. \ref{fig:fig11}. At particle hole symmetry ($\tilde\epsilon_d=0$) the influence of the contact asymmetry vanishes, except for $c_{\theta V}$. A more detailed analysis of this effect is presented in Fig. \ref{fig:fig11b}, where we show $c_V$ as function of $\Gamma_\mathrm{L}/\Gamma_\mathrm{R}$ for different value of $\tilde\epsilon_d$ and $U$. It can be seen that the overall variations of $c_V$ are rather large, and only as the system goes into the strongly interacting regime (large $U$) the effect of contact asymmetry becomes small,
and it completely vanishes for very large $U$ and $\tilde\epsilon_d=0$, where it reaches the limiting value of 3/2 discussed above. Note that around $\Gamma_\mathrm{L}/\Gamma_\mathrm{R}=1$ (symmetric coupling) $c_V$ varies quadratically for small variations of $\Gamma_\mathrm{L}/\Gamma_\mathrm{R}$ around 1 (see also ESI Sec. S9).

The rSPT provides a consistent description of the low-temperature, low-field, low-bias transport properties of the Anderson model. When the parameters are calculated from DFT+NEGF, and combined with NRG and/or Bethe ansatz, the method allows for an effectively first principles calculation of all the transport parameters. If the atomic structure is well defined, as is the case in STM experiment of molecules or other adsorbates on flat surfaces \cite{Droghetti.17}, the approach is predictive on a quantitative level. When the structure is not known, as is the case for the PTM/Au system considered here, the approach allows to estimate ranges of possible electronic coupling coefficients, interactions energies and deviations from the particle-hole asymmetry. In this case the results give a qualitative guidance to experiments as to which atomic structures are expected to lead to Kondo physics in a measurement.

\section{Conclusions}
\label{sec:concl}

The theoretical modeling of Kondo physics in nano-scale devices is usually limited to fitting the parameters of a SIAM to conductivity measurements.
Due to this adjustment of the parameters to the experiment such an approach is therefore not predictive, and the question whether it captures the right physics for a given experiment is therefore open.
Moreover, it does not provide any information on the relationship between the device structure and its conductance as well as its electronic interactions.
In order to overcome this limitation and provide a predictive model here we present a scheme that obtains the required parameters of the SIAM from DFT calculations for realistic atomic structures. Importantly, conductance measurements are inherently a non-equilibrium process, and our novel scheme combining DFT, NEGF, NRG and rSPT is designed to capture such effects.
We derive the equations that relate the equilibrium density of states to the non-equilibrium conductance versus voltage curves, which is necessary to interpret experimental conductance measurements in terms of the electronic and atomic structure of the system.
With this approach it is therefore possible
to calculate the electronic and non-equilibrium transport properties of strongly correlated molecular junctions in a systematic and predictive way effectively from first principles.

We employ the method for the description of the recently measured Au/PTM/Au break-junctions. The main limitation of break-junction experiments is that the statistical nature of the measurements does not allow a direct understanding of the atomic structures responsible for the conductance and its variations. First-principles calculations are therefore essential to gain a full atomistic insight on the system properties. While 
state of the art DFT+NEGF can only be applied to weakly correlated systems, the method presented here is proven to overcome this limitation. In fact, for the Au/PTM/Au break-junction we show how the molecule-electrode contacts affect the energy level alignment, charge transfer, hybridization and, ultimately, the Kondo temperature and conductance.
Importantly, we show that while the Kondo temperature depends only on the total hybridization of the molecules with the electrodes, the experimental conductance depends also on the relative coupling to left and right electrodes, since those determine the current flow. Our projection scheme allows us to obtain these required individual electronic couplings from DFT, and with these we are able to evaluate the low bias conductance versus voltage curves by means of the rSPT.
For PTM molecules weakly coupled to the electrodes, as is the case for a molecule on an idealized perfectly flat Au surface, we predict the Kondo temperature to lie below 
the experimentally accessible limit. In contrast, for asymmetric junctions with molecules on a corrugated Au surface, where the central carbon atom has a good electronic contact with the Au, the calculated Kondo temperature is in good agreement with experiments. 
These results are consistent with the experimental 
finding, where only a limited number of junctions exhibit Kondo features in the conductance at the accessible low temperatures.

Finally, we note that for experimental setups, where the atomic structure is well characterized, such as for certain adsorbates or defects on flat metal surfaces, the method will enable quantitative comparisons with low-noise experiments. By eliminating free parameters it can therefore lead to a systematic understanding of the non-equilibrium Kondo physics of molecular systems.
The inclusion of the rSPT allows to predict systematic changes in non-linear transport at low voltage, temperature and magnetic field, which cannot be addressed directly from state of the art calculations of the transmission coefficient alone. Such changes can be induced experimentally, for example by varying the scanning tip height, which modifies the asymmetry in the electronic coupling to the electrodes, and these can then be calculated effectively from first principles with the approach presented here.
Our method therefore paves the way toward the rational design of Kondo systems, and the possibility of performing systematic comparisons with unprecedented accuracy between theory and experiments.

\section*{Conflicts of interest}
There are no conflicts to declare.

\section*{Acknowledgements}
A.D. and I.R. would like to thank E. Burzur\'i and A.V. Rudnev for useful discussions about transport experiments in Au/PTM/Au junctions, and C. Weber for insightful discussions on the Kondo physics. A.D. and I.R. acknowledge the financial support from the EU project ACMOL (FET Young Explorers, No. 618082). A.D. received additional support from EU Marie Sklodowska-Curie project SPINMAN (No. SEP-210189940) and from the Ministerio de Econom\'ia y Competitividad de Espa\~na (No. FPDI-2013-16641). I. R. acknowledges additional financial support from the
EU H2020 programme PETMEM project (Grant No. 688282). I.R. thanks the Cambridge CSD3 HPC centre for providing part of the computing resources. 
W.H.A., L.C. and D.V. acknowledge the financial support from the Deutsche Forschungsgemeinschaft through TRR80/F6,  TRR80/G7 and the FOR1346/P3.  
E.~Mu\~noz acknowledges financial support by Fondecyt (Chile) No. 1141146. M.M.Radonji\'c acknowledges the support from Ministry of Education, Science, and Technological Development of the Republic of Serbia under project ON171017.
S.~Kirchner acknowledges support by the National Key R\&D Program of the MOST of China, grant No.\ 2016YFA0300202, the National Science Foundation of China, grant No.\ 11774307 and No.\ 11474250, and the U.S. Army RDECOM - Atlantic Grant No. W911NF-17-1-0108.


\providecommand*{\mcitethebibliography}{\thebibliography}
\csname @ifundefined\endcsname{endmcitethebibliography}
{\let\endmcitethebibliography\endthebibliography}{}

\clearpage

\title{Supplementary Information:\\Predicting the conductance of strongly correlated molecules:  the Kondo effect in perchlorotriphenylmethyl/Au junctions}
\author{W. H. Appelt$^{a,b}$, A. Droghetti$^c$, L. Chioncel$^{b,d}$, M. M. Radonji\'c$^e$, E. Mu\~noz$^f$, S. Kirchner$^g$, D. Vollhardt$^{d}$, I. Rungger$^{h,*}$}

\affil{$^a$Theoretical Physics II, Institute of Physics, University of Augsburg, D-86135 Augsburg, Germany}
\affil{$^{b}$Augsburg Center for Innovative Technologies, University of Augsburg, D-86135 Augsburg, Germany}
\affil{$^{c}$Nano-Bio Spectroscopy Group and European Theoretical Spectroscopy Facility (ETSF), Centro de Fisica de Materiales, Universidad del Pais Vasco, Avenida Tolosa 72, 20018 San Sebastian, Spain}
\affil{$^{d}$Theoretical Physics III, Center for Electronic
Correlations and Magnetism, Institute of Physics, University of
Augsburg, D-86135 Augsburg, Germany}
\affil{$^{e}$Scientific Computing Laboratory, Center for the Study of Complex Systems, Institute of Physics Belgrade, University of Belgrade, Pregrevica 118, 11080 Belgrade, Serbia}
\affil{$^{f}$Facultad de Fisica, Pontificia Universidad Cat\'olica de Chile, Casilla 306, Santiago 22, Chile}
\affil{$^g$ Zhejiang Institute of Modern Physics, Zhejiang University, Hangzhou, Zhejiang 310027, China}
\affil{$^h$ National Physical Laboratory, Teddington, TW11 0LW, United Kingdom}
\affil{$^*$ E-mail: ivan.rungger@npl.co.uk}
\date{}

\maketitle

\beginsupplement

\section{Computational details of the DFT and DFT+NEGF calculations}
\label{app:DFTdetails} 

The DFT calculations in Sec.~2
of the main manuscript are performed with the pseudo-potential code SIESTA 
\cite{Siesta_esi} and the all-electron code FHI-aims \cite{AIMS_esi, Ren_esi,Caruso_esi}. The computational details are the same as those used in previous works \cite{Droghetti.17_esi,AndreaGW_esi}. 
For the gas phase molecule we use the Perdew-Burke-Ernzerhof (PBE) generalized gradient approximation (GGA) \cite{pbe1_esi,pbe2_esi} of the exchange-correlation functional, as well as the hybrid functional PBE0 \cite{Adamo_esi}. 
Additionally, the energy gaps in the density of states (DOS) are also computed with the spin-polarized $G_0W_0$@PBE0 approximation of the many-body perturbation theory implemented in FHI-aims, and the results are presented in the Sec. \ref{app:DFTspin}.
Geometry optimizations for the molecule integrated into the Au electrodes are carried out with FHI-aims in a supercell approach and by employing the functional PBE+vdW$^{surf}$ \cite{Tkatchenko_esi,Ruiz_esi}. 
The molecule and all the Au atoms, except for those at the boundary and those directly connected to leads in the transport setup, are allowed to relax until forces are smaller than $0.01$ eV/\AA.

DFT+NEGF calculations are performed with the \textit{Smeagol} code \cite{smeagol_esi,ivan_esi}. The geometries are obtained by joining the central part of the scattering region (optimized by using FHI-AIMS) to the Au electrodes, and all the considered structures are shown in Fig.~1 of the main manuscript.
The simulation parameters for the DFT+NEGF calculations correspond to those outlined in Ref. \citenum{Droghetti.17_esi}. We employ non-spin-polarized DFT calculations, since the formation of either the local magnetic moment or the Kondo state is accounted for subsequently in the DFT+NEGF+NRG solutions. Results for a set of spin-polarized DFT+NEGF calculations are shown in Sec. \ref{app:DFTspin}.

\section{Spin-polarized DFT and DFT+NEGF calculations}
\label{app:DFTspin}
\begin{figure}
	\centering
	\includegraphics[width=0.4\textwidth,clip=true]{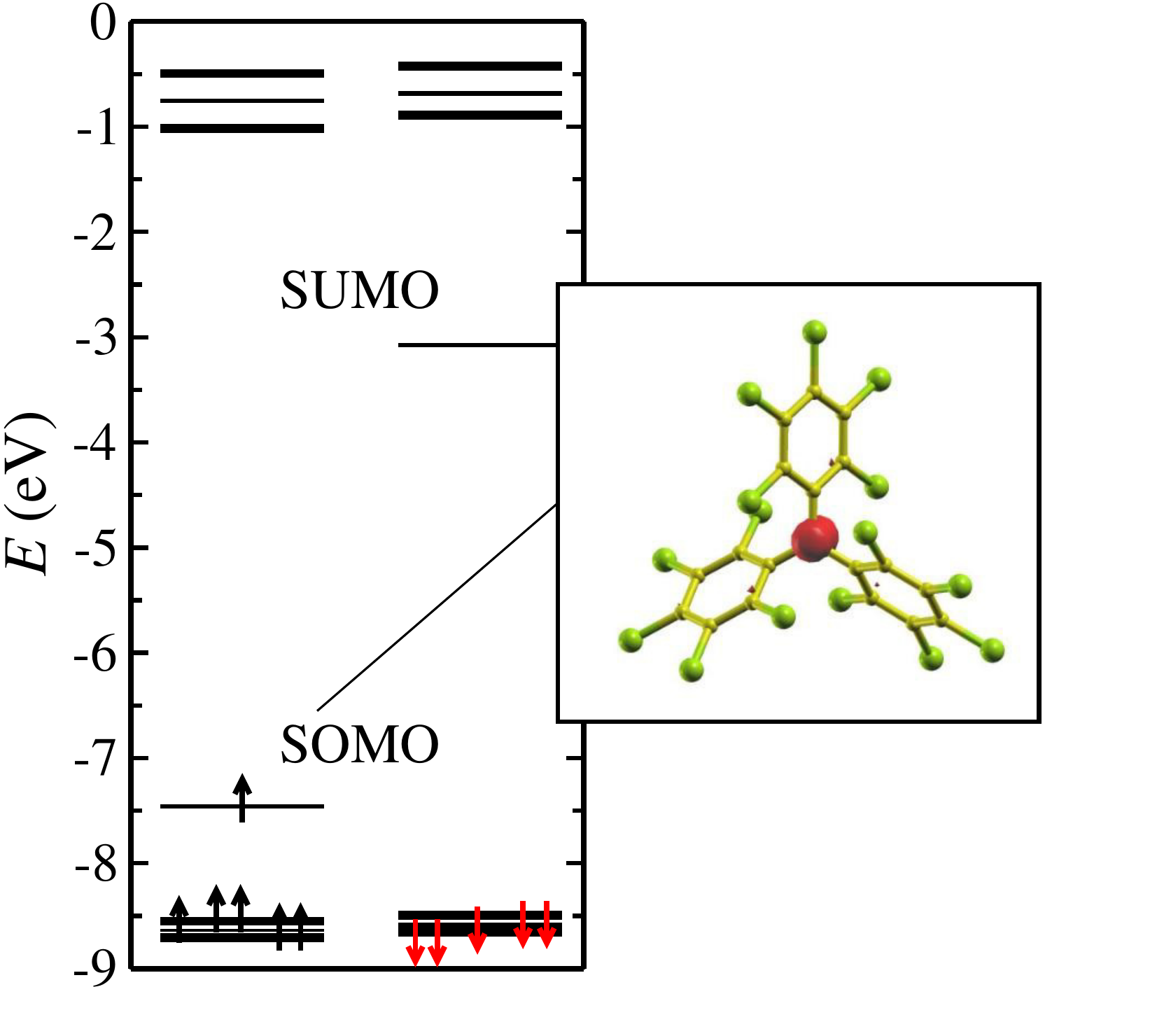}
\caption{Highest occupied and lowest unoccupied molecular orbitals of the PTM radical computed with G$_0$W$_0$@PBE0. The SOMO and the SUMO are explicitly indicated. The SOMO-SUMO gap represents the gas phase charging energy $U$. Inset: charge isosurface for the SOMO (orange bubble); C atoms are in yellow, while Cl atoms are in light green. }
	\label{fig:radical}
\end{figure}

The energy spectrum for the isolated PTM molecule for spin-polarized electrons
is shown in Fig.~\ref{fig:radical}. The calculations are performed by using $G_0W_0$@PBE0, which usually gives an accurate description of the energy levels of small molecules~\cite{Marom_esi}. It can be seen that all spin-up and spin-down levels are occupied up to about $-8.5$ eV. At $-7.5$ eV there is an additional spin-up occupied state, the SOMO. The corresponding spin-down state is therefore empty, and its energy is separated by the charging energy $U=4.31$ eV from that of the SOMO. This $U$ value is in quite close agreement to the values obtained via finite energy difference calculations given in Sec.~2
of the main manuscript. 
Within the present spin-polarized picture, this empty down-spin state is usually called the singly unoccupied molecular orbital (SUMO). For energies above this state the spectrum is again approximately degenerate for up- and down-spins.\\

\begin{figure}
	\centering
	\includegraphics[width=0.45\textwidth,clip=true]{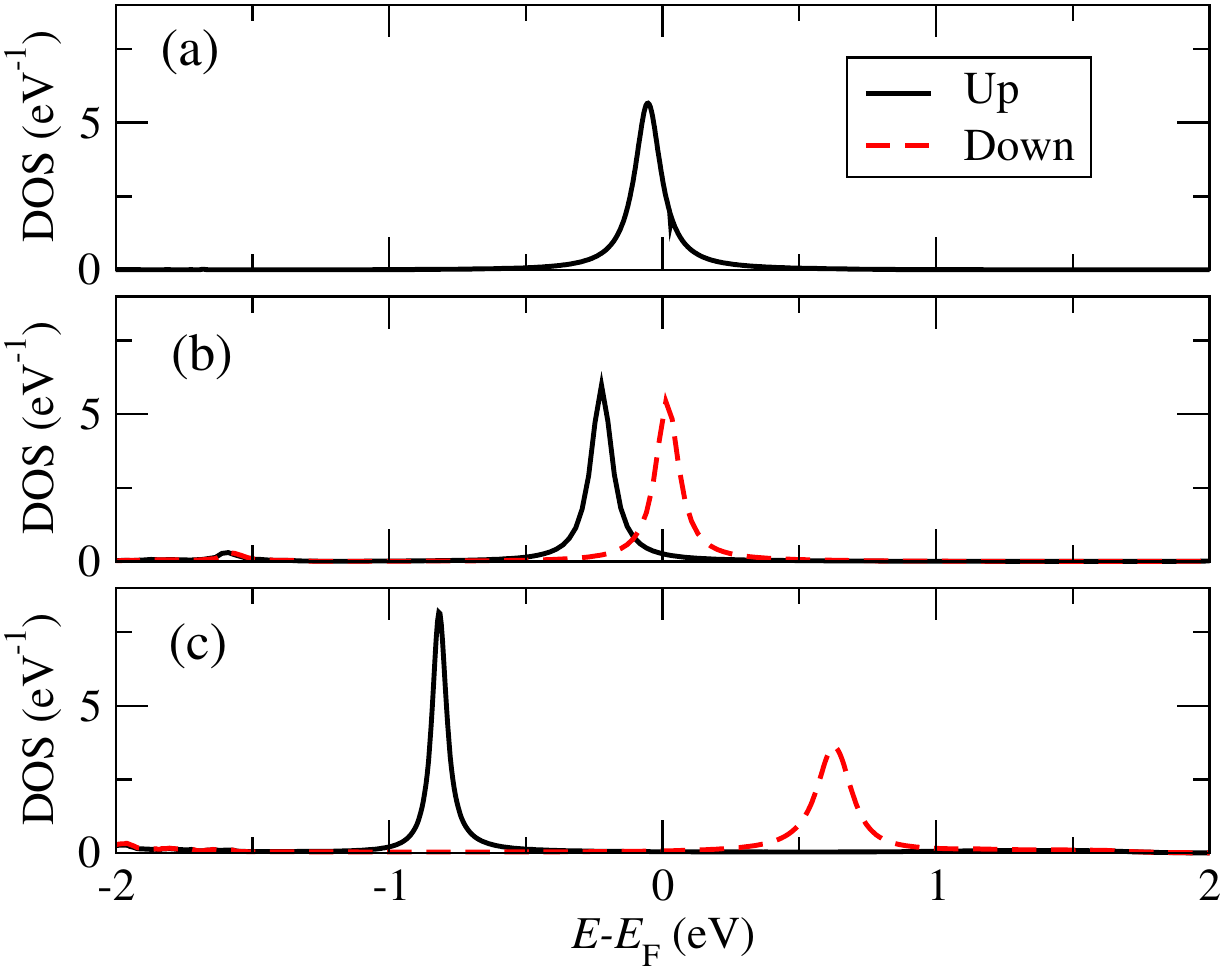}
\caption{DOS of the PTM radical on Au for configuration B2 obtained within a) non-spin-polarized LDA, b) spin-polarized LDA,  and c) spin-polarized LDA+SCO, with a correction energy of 1 eV.}
	\label{fig:dossco}
\end{figure}
For the molecule adsorbed on the Au substrate we use DFT within the local spin-density approximation (LSDA) to compute the density of states (DOS), since G$_0$W$_0$@PBE0 becomes computationally too demanding. However, the LSDA DOS of the molecule suffers from the well-known DFT Kohn-Sham (KS) gap error. 
On the one hand, the KS SOMO-SUMO gap for the gas phase molecule is only about 0.4 eV, and 
is therefore significantly smaller than the $G_0W_0$@PBE0 result. This may lead to a drastic overestimation of the computed conductance.
On the other hand, the SOMO-SUMO gap for molecules adsorbed on a metal substrate is known to shrink with respect to that in the gas phase due to the image charge effect~\cite{Perrin_esi,Lu_esi}. This effect is not captured by the KS description of the energy levels spectrum with local and semi-local functionals~\cite{Neaton_esi,Lastra_esi,Amaury_esi}. A fortuitous error cancellation between the gap underestimation and the neglect of image-charge effects 
 may sometimes happen, but this is not generally the case and corrections are 
 therefore needed.

A reliable estimate of the gap for adsorbed molecules is generally obtained from the gas phase value computed with $G_0W_0$, to which one then adds a classical image-charge correction~\cite{Neaton_esi,Lastra_esi,Amaury_esi}. 
The practical way to adjust the gap in KS DFT-based electron transport calculations is then to add a scissor operator (SCO) correction to the KS eigenvalues~\cite{Quek_esi,Strange_esi,Amaury2_esi,DroghettiNatComm_esi}. 
In Fig.~\ref{fig:dossco} (b)-(c) the DOS of the adsorbed molecule obtained without and with applying a scissor operator correction of 1 eV is shown, and compared also to the non-spin-polarized calculation in Fig.~\ref{fig:dossco}(a).

Note that while the SCO correction can move the SOMO and SUMO peaks to the correct 
energies, the DOS does not show any Kondo feature at $E_\mathrm{F}$. Furthermore,  magnetism in these DFT calculations is described by breaking the symmetry via the unequal occupation of spin up and spin down states~\cite{Kepenekian_esi}. This does not correspond to the correct many-body picture, where the spin symmetry is preserved unless there is an external Zeeman field. 
The use of KS-DFT with the SCO in transport is effectively equivalent to solving the SIAM in Eq.~(1)
with a static mean-field approximation~\cite{Droghetti.17_esi,Cehovin_esi}. 
To obtain the correct Kondo physics and magnetic behavior it is necessary to add many-body corrections to the KS-DFT, for which we use the NRG in Sec. 3.3 of the main manuscript.

\section{Kondo temperature}
\label{sec:app_tk}

The Kondo model is defined by the Hamiltonian
\begin{equation}
H_\mathrm{K}= H_d + \frac{1}{2}\;J \vec{S}_d \sum_{k,k',\sigma,\sigma'} c_{k,\sigma}^\dagger \vec{\sigma}_{\sigma \sigma'} c_{k',\sigma'}
\end{equation}
where $\vec{S}_d=\frac{1}{2} \sum_{\sigma\sigma'}d^\dagger_{\sigma} \vec{\sigma}_{\sigma\sigma'} d_{\sigma'}$ is the spin operator of the impurity site, the vector $\vec{\sigma}$ contains the Pauli-matrices as components, and $J$ is a number representing an effective exchange coupling.
As in Sec.~3.1
of the main manuscript, $c_{k,\sigma}$ ($c_{k,\sigma}^\dagger$) and $d_{\sigma}$ ($d^\dagger_{\sigma}$) are the fermionic annihilation (creation) operators for the bath electrons and the impurity electrons, respectively. 
The SIAM in Eq.~(1)
maps onto the Kondo model under the following conditions.
For a k-independent  hybridization matrix element, $V_k=V$, a constant DOS of the bath, $\rho$, and therefore a constant $\Gamma= 2\pi V^2 \rho$,
the impurity spin $\vec{S}_d$ is coupled via $J=2V^2 \left(\frac{1}{\epsilon_d}-\frac{1}{\epsilon_d +U}  \right)$ to the bath~\cite{schrieffer1966relation_esi}. Note that the strong correlation limit ($U\gg\Gamma$) for the SIAM corresponds to the weakly coupled case in the Kondo model $J\rho \ll 1$.  

A method to approach the so-called Kondo problem without employing perturbation theory or 
mean-field decoupling techniques was developed in a series of papers by Anderson and coworkers \cite{PhysRev.164.352_esi,PhysRevLett.23.89_esi,PhysRevB.1.4464_esi,PhysRevB.1.1522_esi}, which then led to the renormalization group approach (RG) by Wilson~\cite{RevModPhys.47.773_esi}. 
In this method it became apparent that the generation of the low energy scale signals the renormalization group flow to a strong-coupling fixed point and a formation of a singlet state for temperatures $\theta\ll \theta_\mathrm{L}$, where
\begin{equation}\label{T_K_Jrho}
\theta_\mathrm{L} = \frac{1}{2}\sqrt{\Gamma U} e^{-\frac{1}{J\rho}}.
\end{equation}
We can replace $J$ and $\rho$ by $\epsilon$ and $\Gamma$ to obtain the equivalent equation (Ref. \citenum{Hewson.book_esi}, page 168)
\begin{equation} 
\theta_\mathrm{L} = \frac{1}{2}\sqrt{\Gamma U} e^{\frac{\pi \epsilon_d  \left(\epsilon_d+U \right)}{U \Gamma}}.\label{gamma_u_t_asym2}
\end{equation} 

Since $\theta_\mathrm{L}$ is defined by $\chi_s(\theta=0)=(g\mu_\mathrm{B})^2/4k_\mathrm{B} \theta_\mathrm{L}$, 
it can be determined from the low-temperature limit of the magnetic susceptibility (see also page 155 in Ref. \citenum{Hewson.book_esi}). 
For the half-filling case, where $\epsilon_d=-U/2$, the temperature $\theta_\mathrm{L}$ agrees with Wilson's numerical result up to a constant (the Wilson number)~\cite{an.lo.81_esi}
\begin{equation}
c_\mathrm{W}=\frac{\theta_\mathrm{W}}{\theta_\mathrm{L}}=0.41072.
\end{equation}
Here $\theta_\mathrm{W}$ denotes the Kondo temperature deduced from the renormalization group calculations by Krishna-murthy \textit{et al.} ~\cite{PhysRevB.21.1003_esi}, which is equal to $\theta_\mathrm{L}$ up to the scaling factor $c_\mathrm{W}$.

\section{Computational details of the NRG calculations}
\label{sec:app_hyb}

The numerical renormalization group (NRG) method is a very powerful tool for the solution of effective impurity models~\cite{PhysRevB.4.3174_esi,PhysRevB.4.3184_esi,RevModPhys.80.395_esi,he.ta.98_esi,PhysRevB.21.1003_esi,PhysRevB.21.1044_esi}. The NRG is a non-perturbative approach and allows to access 
arbitrarily small energy scales, which is essential for the description of systems with characteristic temperatures of the order of $10\,$K and below.
NRG has been extended to handle arbitrary hybridization functions as input and allows one to produce dynamical quantities such as the impurity self-energy~\cite{bulla1998numerical_esi}. 
In the present article we apply the recent methodological developments by \v{Z}itko~\cite{PhysRevB.84.085142_esi}  in computing dynamical quantities, and we employ the z-averaging technique proposed by Oliveira and Oliveira~\cite{PhysRevB.49.11986_esi}. 
Technical details for the application of the NRG to the Anderson model have been described in Ref.~\citenum{RevModPhys.80.395_esi}.

Here we only give a brief overview of the key steps to setup the NRG procedure. First, we divide the energy range of the bath spectral function into a set of logarithmic intervals, hence reducing the continuous spectrum to a discrete set of states (logarithmic discretization).
The discretized version of the model Hamiltonian is mapped onto a one-dimensional system consisting of a semi-infinite chain of sites.
The magnetic impurity, which is left unchanged by the unitary transformation, constitutes the first site in the semi-infinite chain.
Only the coupling to the bath degrees of freedoms and the bath-geometry are modified. 
An iterative diagonalization of the impurity site, coupled to its adjacent bath site, is performed, where high energy levels are truncated.
The truncation error can be controlled as long as the Hamiltonian parameters, like the on-site energies and couplings of adjacent sites along the chain, are well separated in energy.
The separation of energy scales is guaranteed by the logarithmic discretization. 
The iterative diagonalization together with the rescaling of energies is understood as a mapping of the initial impurity site to an effective impurity site coupled to a bath with one bath site removed (the RG-step).
Static thermodynamic and dynamic quantities can both be calculated from the knowledge of the energy spectrum and the eigenstates of the system.
Care has to be taken, however, when one tries to analyze dynamical quantities, since the energy resolution is affected to some extend by the logarithmic discretization~\cite{PhysRevB.84.085142_esi,PhysRevB.76.165112_esi}.

When temperature dependent quantities are computed in NRG, they are usually affected by truncation errors, since only a finite number of energy eigenvalues can be kept at each RG iteration. However, for thermodynamic quantities such as the impurity entropy, the specific heat and the susceptibility, one can show that the statistical weight of the truncated states is suppressed by the Boltzmann factor~\cite{RevModPhys.47.773_esi,PhysRevB.21.1003_esi,RevModPhys.80.395_esi,PhysRevB.49.11986_esi}. We therefore use this scheme in order to compute the temperature dependence of the static susceptibility in Fig.~5 of the main manuscript.
For dynamical quantities, such as the impurity spectral function, the situation is more complicated, since here the information of all energy scales enters. More elaborate schemes need to be employed in this case (see for example Ref. \citenum{RevModPhys.80.395_esi} and references therein).

\begin{figure}
	\centering
  \includegraphics[width=0.46\textwidth,clip=true]{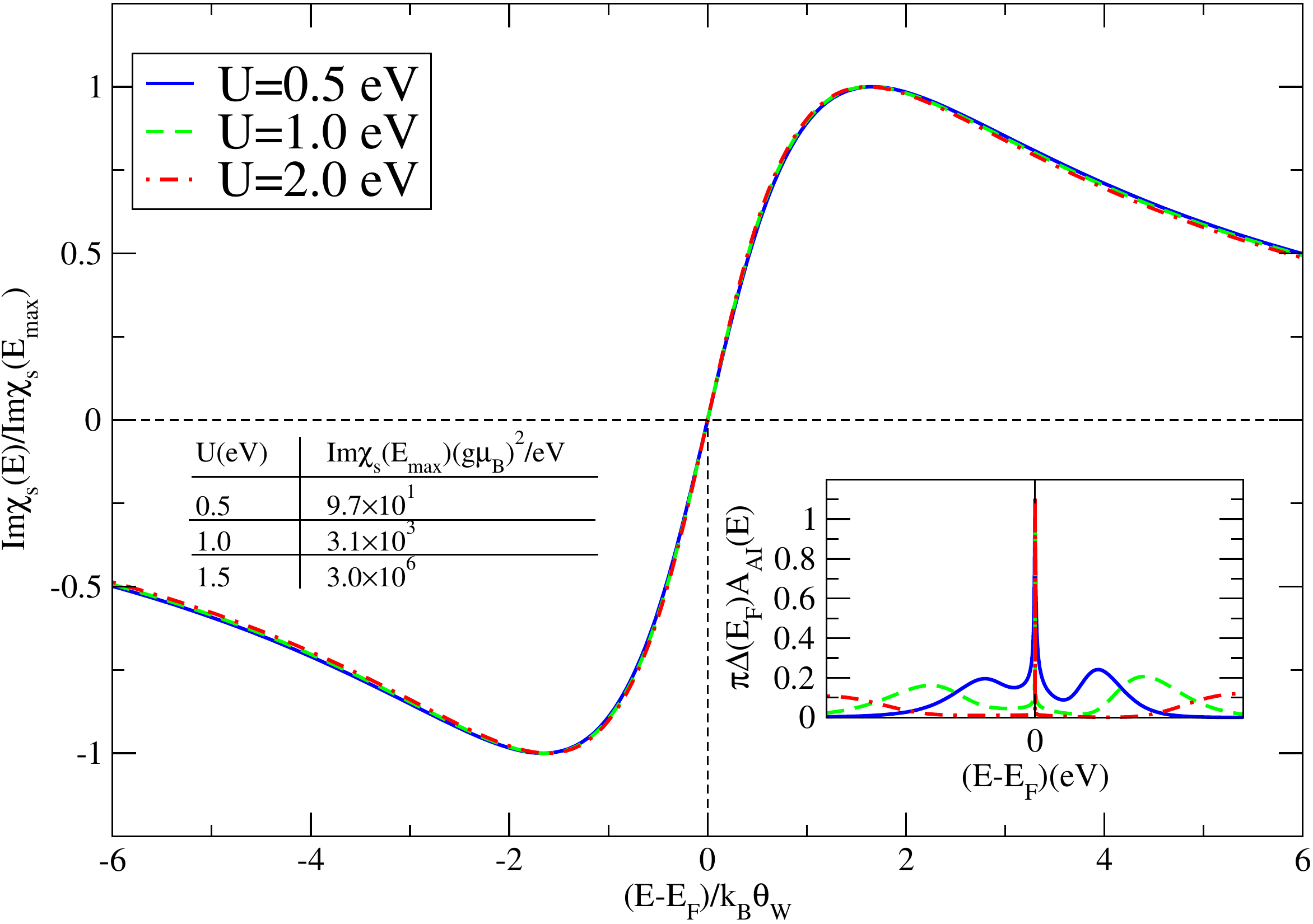}
\caption{Dynamical spin susceptibility $\chi_s(E,\theta=0)$ for different interaction strengths, calculated for hybridization function presented in Fig.~
4 of the main manuscript,
for configuration B4 (see Fig.~
1 of the main manuscript),
and $\theta=0\,$K. Inset: impurity density of states $A_{\mathrm{AI}}(E)$. The calculation is performed close to the half-filled case. 
}
	\label{fig:fig91}
\end{figure}
We introduce an energy cutoff for the hybridization function.
The left and right energy cutoffs are $[-1.20,0.91]\,$eV, $[-1.33,0.30]\,$eV, and $[-0.35,0.80]\,$eV for the B2, B4, and T4 configurations, respectively.
From the truncated imaginary part we compute the real part by performing a Hilbert transformation.
This procedure ensures that the analytic properties of $\Delta(\omega)$ are not modified due to the discretization and truncation.
For large $U$ it can be expected that the bandwidth of the conduction electrons is irrelevant for the low frequency behavior.
The discretization scheme proposed in Ref. \citenum{zi.pr.09_esi} is used, which corrects the systematic error in the first energy interval. The discretization parameter $\Lambda=2$ is applied, and $5000$ states are kept at each RG step.

\section{Impurity susceptibility}
\label{app:susc}
Here we derive the impurity contribution to the isothermal magnetic susceptibility, which is needed for accessing $\theta_\mathrm{W}$. 
It is closely related to the Matsubara susceptibility $\chi^s_n(\beta)$, which is defined as (paramagnetic case):
\begin{equation}
\chi^s_n(\beta)= \int_0^\beta d\tau  e^{i\omega_n \tau}\langle T_\tau S_\mathrm{AI}^z(\tau)S_\mathrm{AI}^z(0) \rangle.
\label{eq:suscept_temp}
\end{equation}
Here $S_\text{AI}^z(\tau)$ is the spin operator on the impurity site $S^z_\text{AI}=\frac{1}{2}(d_\uparrow^{\dagger}d_\uparrow - d_\downarrow^{\dagger}d_\downarrow)$ in the Heisenberg representation for imaginary times $\tau$.
We define the inverse temperature $\beta=1/k_\mathrm{B}\theta$, and denote with $T_\tau$ the time ordering operator, which moves the earlier times to the right.
The brackets $\langle O\rangle$ on an operator $ O$ denote the thermodynamic average $\langle  O\rangle=\text{Tr}\left[e^{-\beta ( H-\mu  N-\Omega)}  O\right]$, with $e^{-\beta \Omega}=\text{Tr}\left[e^{-\beta( H-\mu N)} \right]$, where $\Omega$ is the grand canonical potential, $\mu$ is the chemical potential, and $N$ is the particle number operator.
We do not distinguish the Fermi energy $E_\mathrm{F}$ from the chemical potential $\mu$ in this article, and hence both are employed.
The use of $E_\mathrm{F}$ is commonly used in the context of the first principles DFT-based calculations, while
the chemical potential is favored in the context of statistical physics.
The Matsubara frequencies are the complex energies $i\omega_n=i2n\pi /\beta$, where $n=0,1,2,\dots $.
The index $n$ in Eq.~\ref{eq:suscept_temp} corresponds to the $n$-th Matsubara frequency.
The isothermal magnetic susceptibility is given by the Matsubara susceptibility at the first Matsubara point $\chi^s_{n=0}(\beta)$.
In the following we show how we calculate the isothermal magnetic susceptibility for finite temperatures. 
The total spin, projected on the $z$-direction, commutes with the Hamiltonian ($\langle S_\mathrm{tot}^z(\tau)S_\mathrm{tot}^z(0)\rangle=\langle \left(S_\mathrm{tot}^z\right)^2 \rangle$) so that we find:
\begin{eqnarray}
\chi^\text{tot}_n(\beta)&=& \int_0^\beta d\tau e^{i\omega_n\tau}\langle T_\tau S_\mathrm{tot}^z(\tau)S_\mathrm{tot}^z(0) \rangle\\
\chi^\text{tot}_{n=0}(\beta)&=& \int_0^\beta d\tau \langle T_\tau S_\mathrm{tot}^z(\tau)S_\mathrm{tot}^z(0) \rangle\\
&=&\beta \langle \left(S_\mathrm{tot}^z\right)^2 \rangle,
\label{eq:sz_tot}
\end{eqnarray}
where $S_\mathrm{tot}^z=\frac{1}{2}(d^\dagger_\uparrow d_\uparrow-d^\dagger_\downarrow d_\downarrow)+\frac{1}{2}\sum_{k} (c^\dagger_{k,\uparrow} d_{k,\uparrow}-d^\dagger_{k,\downarrow} d_{k,\downarrow}) $.
From Eq.~(\ref{eq:sz_tot}) one subtracts the susceptibility of a reference system, i.e. of
the system without impurity $\chi^{(0),\text{tot}}_{n=0}(\beta)$.
This gives Wilson's definition of the impurity contribution to the susceptibility~\cite{RevModPhys.80.395_esi,RevModPhys.47.773_esi}:
\begin{equation}
\chi_s(\beta) \equiv \chi^s_{n=0}(\beta)=\chi^\text{tot}_{n=0}(\beta)-\chi^{(0),\text{tot}}_{n=0}(\beta).
\end{equation}
The latter is used to determine the Kondo scaling.

The dynamical spin susceptibility is defined as~\cite{mahan2013many_esi}:
\begin{equation}
\chi_s(E,\beta)= i \int_{-\infty}^0 dt e^{iEt} \langle [S_\mathrm{AI}^z(t),S_\mathrm{AI}^z(0)] \rangle,
\label{eq:suscept}
\end{equation}
where $S_\text{AI}^z(t)$ is spin operator on the impurity site, $S^z_\text{AI}=\frac{1}{2}(d_\uparrow^{\dagger}d_\uparrow - d_\downarrow^{\dagger}d_\downarrow)$, in the Heisenberg representation for real times $t$.
The susceptibilities in Eq.~\ref{eq:suscept} and Eq.~\ref{eq:suscept_temp} are related by analytic continuation $i\omega_n \to E+i\eta$. 
In Fig.~\ref{fig:fig91} we show the imaginary part of $\chi_s(E,\theta=0)$ at zero temperature,
and the inset shows the density of states $A_\mathrm{AI}(E)=-\frac{1}{\pi}{\rm Im}G_{d\sigma,d\sigma}(E)$ of the impurity site, where $G_{d\sigma,d\sigma}(E)=-i \int_{-\infty}^0 dt e^{iEt} \langle [d_{\sigma}(t),d^\dagger_{\sigma}(0)] \rangle$ is the one-electron Green's function on the impurity site. A clear three-peak
structure is found which is characteristic for the SIAM.
In particular, the low-energy physics is described by a Kondo resonance in the 
DOS around $E=E_\mathrm{F}=0$, which is gradually suppressed upon increasing $U$ (not shown). The atomic levels, broadened by the coupling to the conducting electrons, are located at $\epsilon_d \pm U/2$. The energy axis of $\mathrm {Im}\chi_s(E,\theta=0)$ is scaled by $\theta_\mathrm{W}$, which is determined from the impurity contribution to the isothermal magnetic susceptibility $\chi_s(0,\theta=0)$ defined above.
One can clearly see that the $\mathrm {Im}\chi_s(E,\theta=0)$ reaches a minimum (maximum)  around $(E-E_\mathrm{F})/\theta_\mathrm{W} \approx 1 $ (around $(E-E_\mathrm{F})/\theta_\mathrm{W} \approx -1$). The vertical axis is scaled by the maximum value of ${\mathrm {Im}}\chi_s(E,\theta=0)$, denoted by $\mathrm{Im}\chi_s(E_\text{max},\theta=0)$, as given in the table in Fig.~\ref{fig:fig91}. 

\section{Full width at half maximum of the Kondo peak}
\label{sec:app_fwhm}
\begin{table*}[t]
\centering
\begin{tabular}{ c | c | c| c| c | c | c | c }
Configuration & $z_\Sigma$ & $\Delta_\Sigma$ (meV) & $\tilde\Delta_\Sigma$ (meV) & $W_\Sigma$ (meV) &$W_\mathrm{Ref}$ (meV) &$\Delta=\frac{\Gamma}{2} (meV)$ & $W_\mathrm{T}$ (meV)  \\
 \hline
B2 & 0.00315 & 67  & 0.210 & 0.381 & 0.599 & 57 & 0.381 \\
B4 & 0.00326 & 67  & 0.215 & 0.395 & 0.612 & 59 & 0.408 \\
T4 & 0.00598 & 75  & 0.449 & 0.816 & 1.265 & 63 & 0.830 \\
\hline
\end{tabular}
\caption{Values of $z_\Sigma$ and $\Delta_\Sigma$ extracted from the low-energy 
many-body self-energy (Eq.~(\ref{eq:sigma_app})), and resulting FWHM $W_\Sigma$ at zero temperature using Eq.~(\ref{eq:fwhm_appendix0}), are compared to the values for $\Delta=\Gamma/2$ obtained directly from the DFT calculations (see Fig.~
1 of the main manuscript)
and of the FWHM obtained from the width of the Kondo peak in the transmission coefficients, $W_\mathrm{T}$. We also present the value obtained from $\tilde\Delta_\Sigma$ using the relation given in Ref. \citenum{PhysRevLett.88.077205_esi}, denoted as $W_\mathrm{Ref}$.}
\label{tab:deltafit_app}
\end{table*}
For a non-spin-polarized system the temperature dependent spin-resolved spectral function $A_\mathrm{AI}=A_\mathrm{AI}^\uparrow=A_\mathrm{AI}^\downarrow$ of the SIAM is given by
\begin{align}
A_\mathrm{AI}(E,\theta)&=\nonumber\\
&\frac{1}{\pi}\frac{\Delta-\mathrm{Im}(\Sigma(E,\theta))}{\left[E-\epsilon_d-\mathrm{Re}(\Sigma(E,\theta))\right]^2+\left[\Delta-\mathrm{Im}(\Sigma(E,\theta))\right]^2},
\label{eq:dose_app}
\end{align}
with $\Delta=\Gamma/2$ and $\epsilon_d$ defined in the main text. The normalization of $A_\mathrm{AI}(E,\theta)$ is chosen in such a way that $\int_{-\infty}^{\infty}A_\mathrm{AI}(E,\theta)dE=1$, so that it is equal to the density of states.
Here $\Delta$ is assumed 
to be energy-independent, and we will discuss the implication of this approximation in the following.
The low-energy behavior of $A_\mathrm{AI}$ is then determined by the low energy 
expansion of the many-body SIAM self-energy $\Sigma(E,\theta)$. In the half-filled, 
strong correlation ($\Delta\ll U$) case considered here we have $\epsilon_d + \mathrm{Re}(\Sigma(0,0))=0$, so that $A_\mathrm{AI}$ is particle-hole symmetric for a constant $\Delta$. 
In this case, the zero voltage low energy expansion of $\Sigma(E,\theta)$ for real energies $E$ is given by \cite{Hewson.93_esi,Munoz.11_esi,Costi.94c_esi}
\begin{equation}
\Sigma(E,\theta)\approx-\epsilon_d+(1-z^{-1}) E -i \frac{\Delta}{2}\left(\frac{E^2}{\tilde\Delta^2}+\frac{\pi^2 k_\mathrm{B}^2\theta^2}{\tilde\Delta^2}\right),
\label{eq:sigma_app}
\end{equation}
with $\tilde{\Delta}=z \Delta=z \Gamma/2$, and $z=(1-\frac{\partial\mathrm{Re}(\Sigma(E))}{\partial E}|_{E=0})^{-1}$, evaluated at zero temperature \cite{Hewson.93_esi,Munoz.11_esi}. We now insert Eq.~(\ref{eq:sigma_app}) into Eq. (\ref{eq:dose_app}), and find that $A_\mathrm{AI}$ is maximal at $E=0$, with the height of the peak given by
\begin{equation}
A_\mathrm{AI}(0,\theta)=\frac{1}{\pi \Delta}\frac{1}{1+\dfrac{\pi^2 k_\mathrm{B}^2\theta^2}{2{\tilde\Delta}^2}}.
\label{eq:dosmax_app}
\end{equation}
The full width at half maximum, $W$, is determined by solving the condition $A_\mathrm{AI}(W/2,\theta)=A_\mathrm{AI}(0,\theta)/2$ for $W$. Using Eqs. (\ref{eq:dose_app}-\ref{eq:dosmax_app}) one then obtains
\begin{equation}
W(\theta,\tilde{\Delta})=
\tilde\Delta 2 \sqrt{2}\sqrt{\sqrt{1+\left(\frac{\pi^2 k_\mathrm{B}^2 \theta^2}{2 \tilde\Delta^2}+1\right)^2}-1}.
\label{eq:fwhm_appendix}
\end{equation}
At zero temperature this FWHM is
\begin{equation}
W(0,\tilde{\Delta})=
\tilde\Delta 2 \sqrt{2} \sqrt{\sqrt{2}-1}.
\label{eq:fwhm_appendix0}
\end{equation}

\begin{figure}
\includegraphics[width=0.4\textwidth,clip=true]{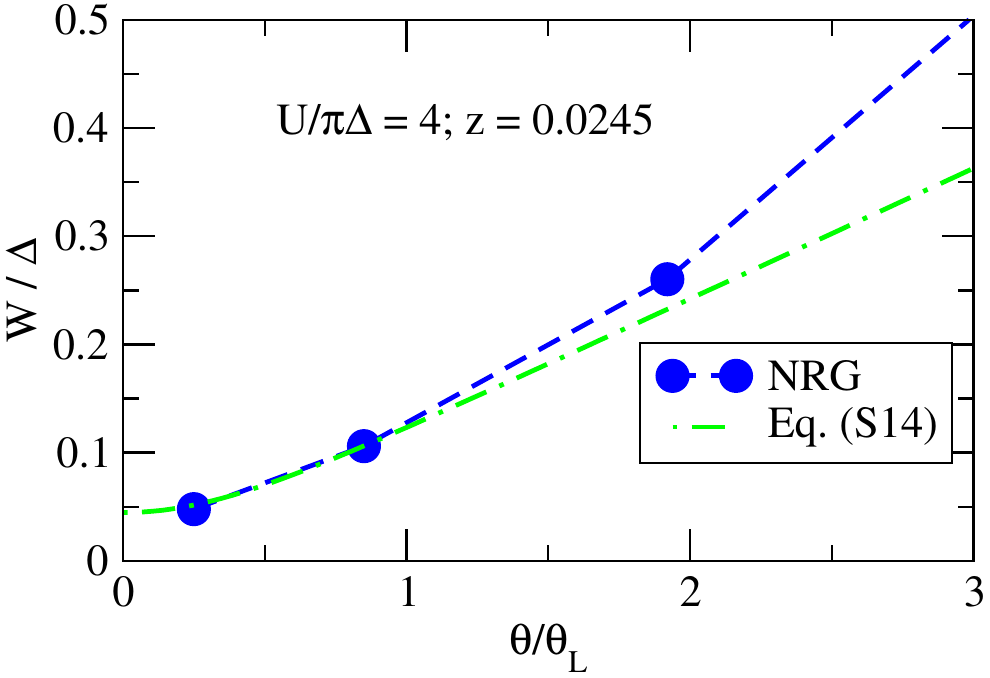}
\caption{Comparison of the results for the full width at half maximum, $W$, of the Kondo peak as function of temperature, obtained by NRG and by the approximated relation in Eq. (\ref{eq:fwhm_appendix}) respectively. The NRG results are extracted from Fig. 6 of Ref. \citenum{Costi.94c_esi}. The parameters used in the NRG calculations are $U/\pi\Delta=4$, and a constant DOS of the electrodes as function of energy is assumed.} \label{fig:fwhm_nrg}
\end{figure}

To estimate the validity of our approximation in Eq. (\ref{eq:fwhm_appendix}) for $W(\theta,\tilde\Delta)$, in Fig. \ref{fig:fwhm_nrg} we compare it to the exact results obtained from NRG calculations in Ref. \citenum{Costi.94c_esi}. The agreement is very good up to about $\theta \approx 2 \theta_\mathrm{L}$, above which the NRG FWHM rises faster. This is mainly due to the fact that in our approximation we assume that there is only the Kondo peak, while in the full spectral function also the two side-peaks at $\pm U/2$ contribute to the spectral function around $E\approx 0$. This contribution increases relative to the height of the Kondo peak as the temperature increases well above $\theta_\mathrm{L}$. 

We can now apply Eq. (\ref{eq:fwhm_appendix}) for the FWHM to our Au-PTM system. We first fit $\Sigma(E,0)$ calculated using the NRG to the expansion in Eq. (\ref{eq:sigma_app}) by adjusting the free parameters in that equation, namely $z$ and $\Delta$ at zero temperature. 
We denote these fitted values as $z_\Sigma$ and $\Delta_\Sigma$ ($\tilde\Delta_\Sigma=z_\Sigma\;\Delta_\Sigma$), and the zero temperature FWHM calculated with these values in Eq. (\ref{eq:fwhm_appendix0}) as $W_\Sigma$. 
The resulting values are given in Table \ref{tab:deltafit_app}, where we also compare them with $\Delta=\Gamma/2$ extracted from the DFT calculations (see Fig. 1 of the main manuscript),
and to the FWHM evaluated directly from the width of the transmission peaks in Fig. 6
of the main manuscript, which we denote as $W_\mathrm{T}$. One can see that $\Delta_\Sigma$ and $\Delta$ are in rather good agreement, confirming the validity of the expansion in Eq. (\ref{eq:sigma_app}) for our system. Since the hybridization function of the Au-PTM system is energy dependent, we can interpret the fitted values $\Delta_\Sigma$ as effective average hybridization strengths. 
In the same way $z_\Sigma$ corresponds to an effective average wave-function renormalization factor, and $\tilde\Delta_\Sigma$ to an effective renormalized hybridization. These therefore correspond to the effective NRG results for $z$ and $\tilde\Delta$, and we report them in Table 1 of the main manuscript.
Importantly, $W_\Sigma$ and $W_\mathrm{T}$ are also in good agreement, confirming the validity of the relations presented in this appendix. We note that in Ref.~\citenum{Nagaoka.02_esi} the zero temperature FWHM is given by a different relation, which we denote as $W_\mathrm{Ref}$, namely $W_\mathrm{Ref}=
2\tilde\Delta\sqrt{2}$. Using our fitted values of $\tilde\Delta_\Sigma$ we also evaluate $W_\mathrm{Ref}$ for all configurations, and the values are reported in Table \ref{tab:deltafit_app}. One can see that $W_\mathrm{Ref}$ systematically overestimates the FWHM when compared to the correct value $W_\mathrm{T}$. This is in contrast to $W_\Sigma$, which agrees well with $W_\mathrm{T}$, showing that our expansion in Eq. (\ref{eq:fwhm_appendix}) provides a more accurate approximation of the Kondo peak width.

\section{Transmission, current and conductance}
\label{sec:app_t}
\begin{figure}
\includegraphics[width=0.48\textwidth,clip=true]{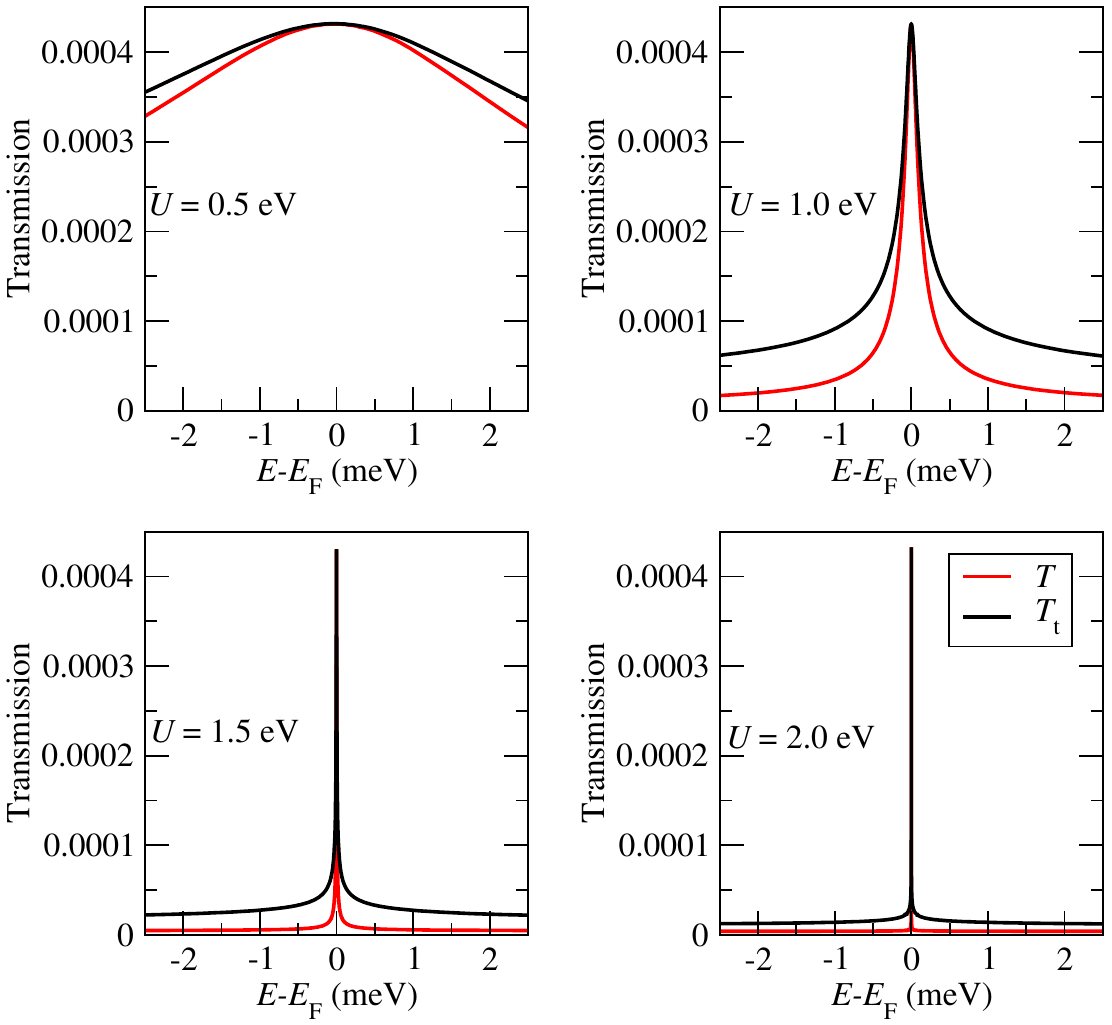}
\caption{Transmission including the NRG self-energy for the B2 structure, for  four
different values of $U$.} \label{fig:TRC_B2_all}
\end{figure}

\begin{figure}
\includegraphics[width=0.48\textwidth,clip=true]{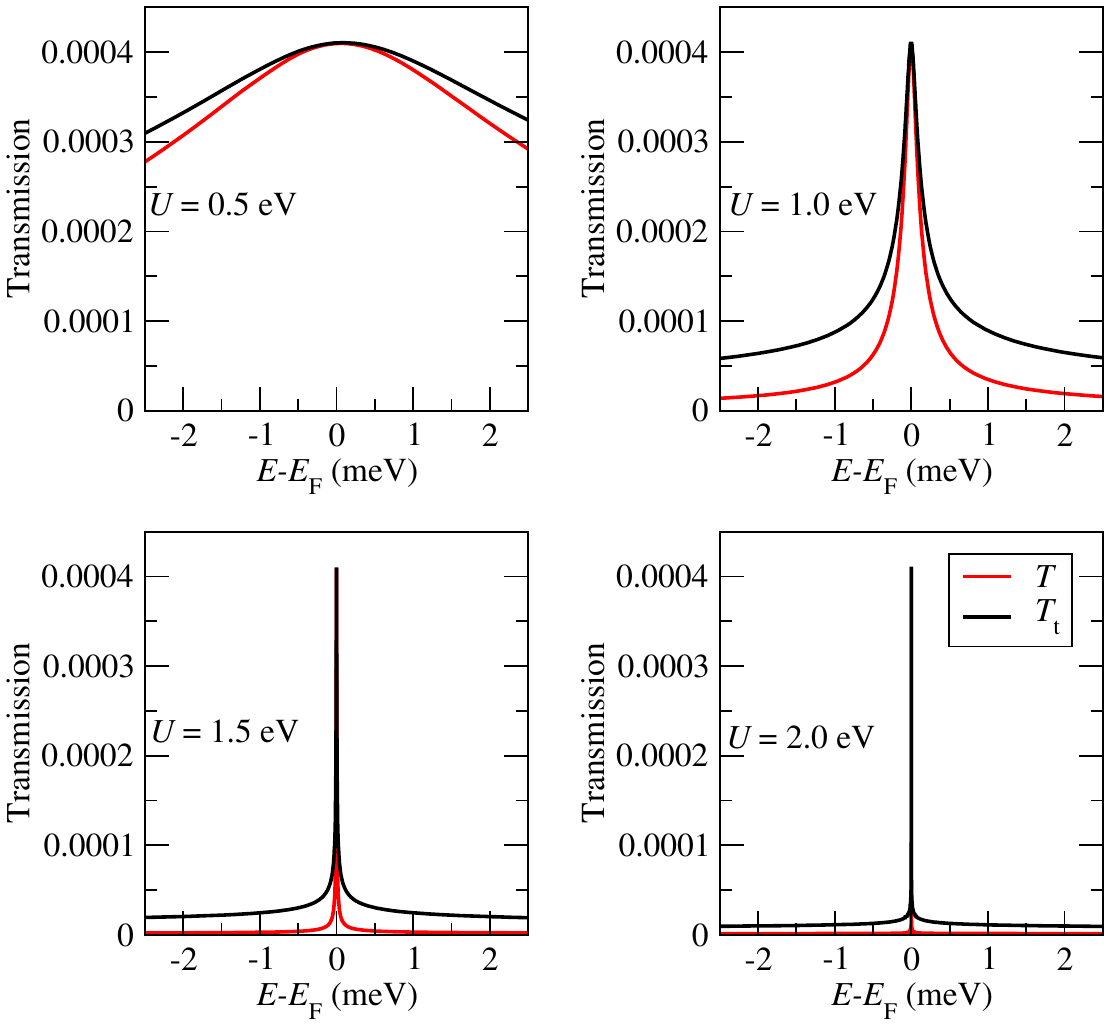}
\caption{Transmission including the NRG self-energy for the B4 structure, for four
different values of $U$.} \label{fig:TRC_B4_all}
\end{figure}

\begin{figure}
\includegraphics[width=0.48\textwidth,clip=true]{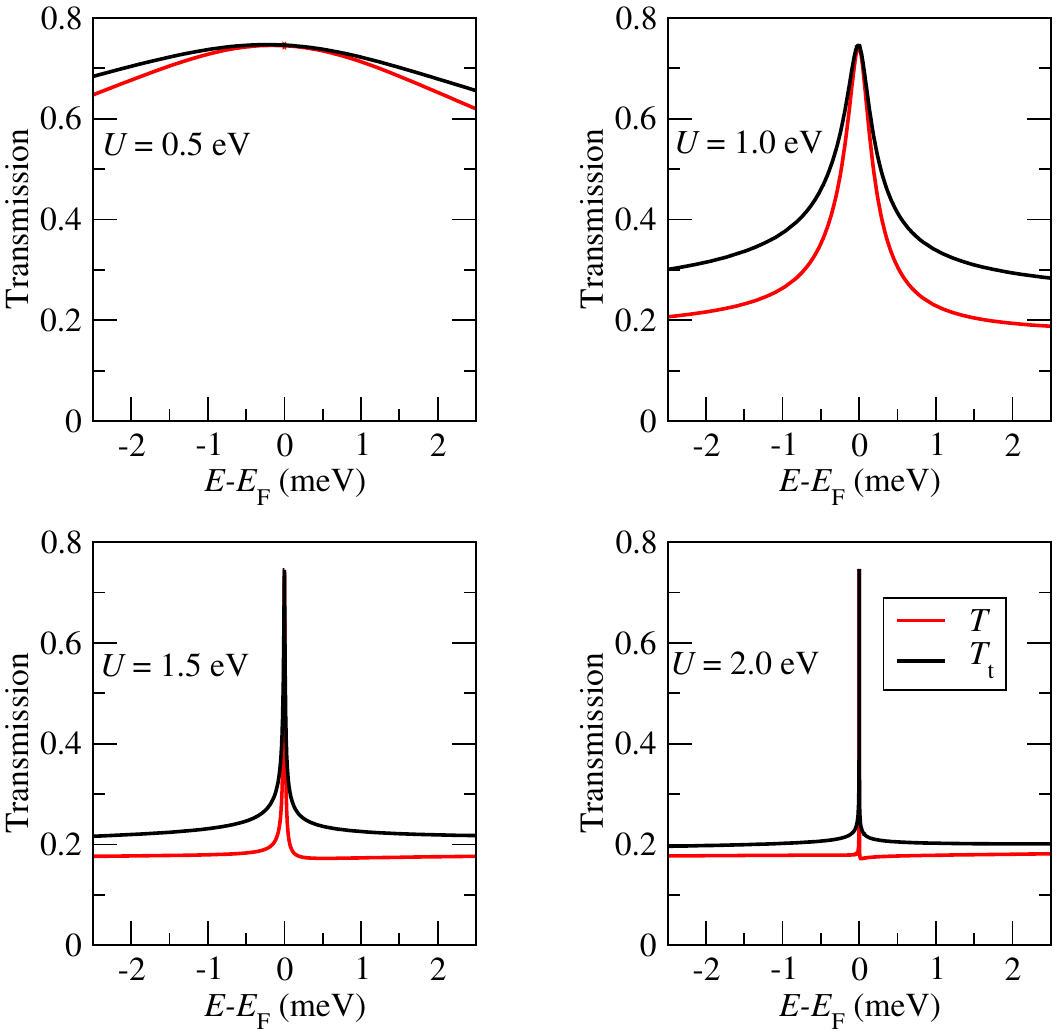}
\caption{Transmission including the NRG self-energy for the T4 structure, for four 
different values of $U$.} \label{fig:TRC_T4_all}
\end{figure}

In the case of the SIAM, the total current can be written as~\cite{Droghetti.17_esi}
\begin{eqnarray}
I=2\frac{e}{h}\int dE \;\left(f_\mathrm{L}(E)-f_\mathrm{R}(E)\right)T_\mathrm{t}(E),
\label{eq:itotal}
\end{eqnarray}
where $f_\mathrm{L(R)}(E)=f(E-\mu_\mathrm{L(R)})$ is the Fermi function of the left (right) electrode with chemical potential $\mu_\mathrm{L(R)}$, $T_\mathrm{t}(E)=T_\mathrm{t}^\uparrow(E)=T^\downarrow_\mathrm{t}(E)$ is the spin-resolved total effective transmission, and the factor 2 takes into account the spin-degeneracy. The applied bias voltage, $V$, is equal to $V=(\mu_\mathrm{L}-\mu_\mathrm{R})/e$, and $T_\mathrm{t}(E)$ is given by
\begin{equation}
T_\mathrm{t}(E)=T_\mathrm{B}(E)+T_\mathrm{AI}(E)+T_\mathrm{I}(E)+T_\mathrm{R,AI}(E),
\label{eq:tedecomptotal}
\end{equation}
and includes elastic and incoherent transmission through the impurity, $T_\mathrm{I}(E)$ and $T_\mathrm{R,AI}(E)$, as well as the background transmission $T_\mathrm{B}(E)$ and the interference term $T_\mathrm{I}(E)$. The total elastic transmission is
$T(E)=T_\mathrm{AI}(E)+T_\mathrm{B}(E)+T_\mathrm{I}(E)$. 
The effective total transmission $T_\mathrm{t,AI}(E)$ through the impurity can be written as~\cite{PhysRevLett.68.2512_esi}
\begin{eqnarray}
T_\mathrm{t,AI}(E)&=&T_\mathrm{AI}(E)+T_\mathrm{R,AI}(E)\nonumber\\
&=&2\pi\frac{\Gamma_\mathrm{L}(E)\Gamma_\mathrm{R}(E)}{\Gamma_\mathrm{L}(E)+
\Gamma_\mathrm{R}(E)}A_\mathrm{AI}(E).
\label{eq:ttai}
\end{eqnarray}
Here it is assumed that $\Gamma_\mathrm{L}(E)=\lambda\,\Gamma_\mathrm{R}(E)$, with $\lambda$ a constant.
In Figs. \ref{fig:TRC_B2_all}-\ref{fig:TRC_T4_all} we present $T(E)$ and $T_\mathrm{t}(E)$ for different values of $U$. The configurations correspond to the ones described in the main text, namely B2, B4, and T4 (see Fig. 6 of the main manuscript).
Note that these are all results for $\theta=0$, so that $T(E_\mathrm{F})=T_\mathrm{t}(E_\mathrm{F})$, and moreover the Kondo peak width becomes vanishingly small as $U$ becomes large.

We can split the total current according to the different contributions in the transmission as
\begin{equation}
I=I_\mathrm{AI}+I_\mathrm{B}+I_\mathrm{I},
\label{eq:app_IExpand}
\end{equation}
where
\begin{eqnarray}
I_\mathrm{AI}&=&\frac{2e}{h}\int dE \;\left(f_\mathrm{L}(E)-f_\mathrm{R}(E)\right)T_\mathrm{t,AI}(E),\label{eq:itotalAI}\\
I_\mathrm{I}&=&2\frac{e}{h}\int dE \;\left(f_\mathrm{L}(E)-f_\mathrm{R}(E)\right)T_\mathrm{I}(E),\\
I_\mathrm{B}&=&2\frac{e}{h}\int dE \;\left(f_\mathrm{L}(E)-f_\mathrm{R}(E)\right)T_\mathrm{B}(E).
\label{eq:ISplitList}
\end{eqnarray}
Note that the transmission coefficients in this set of equations depend also on $\mu_\mathrm{L}$ and $\mu_\mathrm{R}$. In an analogous way we then split the total conductance in its individual contributions
\begin{equation}
G=G_\mathrm{AI}+G_\mathrm{B}+G_\mathrm{I},
\label{eq:app_GExpand}
\end{equation}
with $G_\mathrm{AI}=dI_\mathrm{AI}/dV$, $G_\mathrm{I}=dI_\mathrm{I}/dV$, and $G_\mathrm{B}=dI_\mathrm{B}/dV$.

\section{Conductance of the Kondo resonance}
\label{sec:app_conductance}
\subsection{Low voltage and temperature expansion}
For a finite applied bias voltage the low energy expansion of the many-body self-energy contains additional voltage dependent terms when compared to Eq. (\ref{eq:sigma_app}), as discussed in Ref.~\citenum{Munoz.11_esi}. However, if one considers the highly asymmetric case, where $\Gamma_\mathrm{L} \gg \Gamma_\mathrm{R}$ or $\Gamma_\mathrm{R} \gg \Gamma_\mathrm{L}$, these terms vanish. The SIAM spectral function is then independent of the voltage, except for a rigid shift along the energy axis, which is set by the choice of the arbitrary constant potential shift that can be applied to the whole system. The physical origin of this behaviour can be explained as follows: if either one of $\Gamma_\mathrm{L}$ or $\Gamma_\mathrm{R}$ is very small, then the current through the system is also small, so that the change in the shape of the spectral function with applied voltage is negligible. Without loss of generality we then set the rigid shift of the spectral function to be 0, which is obtained by setting $\mu_\mathrm{L}=E_\mathrm{F}$ and $\mu_\mathrm{R}=E_\mathrm{F}-e V$ when $\Gamma_\mathrm{L} \gg \Gamma_\mathrm{R}$, and by setting $\mu_\mathrm{L}=E_\mathrm{F}+e V$ and $\mu_\mathrm{R}=E_\mathrm{F}$ when $\Gamma_\mathrm{R} \gg \Gamma_\mathrm{L}$. In the following we only consider the case $\Gamma_\mathrm{L} \gg \Gamma_\mathrm{R}$, the equations for $\Gamma_\mathrm{R} \gg \Gamma_\mathrm{L}$ can be obtained in an analogous way.
With this system setup, and with $E_\mathrm{F}$ set to 0, for $\Gamma_\mathrm{L} \gg \Gamma_\mathrm{R}$ the conductance of the AI as function of the applied bias, 
$G_\mathrm{AI}(V,\theta)=dI_\mathrm{AI}/dV$, results from Eq.~(\ref{eq:itotalAI}) to
\begin{equation}
G_\mathrm{AI}(V,\theta) \approx -\frac{2 e^2}{h}\int T_\mathrm{t,AI}(E-eV,\theta) \frac {df(E)}{dE}.\label{eq:app_G}
\end{equation}
At zero temperature $G_\mathrm{AI}(V,\theta=0)\approx(2 e^2/h)T_\mathrm{t,AI}(-e V,\theta=0)$.
At finite low temperatures we can perform a Sommerfeld expansion of the energy integral involved in the calculation of the current, and to second order in $\theta$ the conductance then becomes
\begin{eqnarray}
G_\mathrm{AI}(V,\theta)&\approx&\frac{2 e^2}{h}\left[T_\mathrm{t,AI}(-e V,\theta)+\right.\nonumber\\
&&\left.\frac{\pi^2}{6} \left.\frac{\partial^2 T_\mathrm{t,AI}(E,\theta)}{\partial E^2}\right|_{E=-eV} k_\mathrm{B}^2 \theta^2\right].
\label{eq:gtheta_app}
\end{eqnarray}

Using Eq.~(\ref{eq:ttai}) the temperature dependent $T_\mathrm{t,AI}(E,\theta)$ can be rewritten as $T_\mathrm{t,AI}(E,\theta)=2\pi [\Gamma_\mathrm{L} \Gamma_\mathrm{R}/(\Gamma_\mathrm{L}+\Gamma_\mathrm{R})] A_\mathrm{AI}(E,\theta)$.
With Eqs. (\ref{eq:dose_app}) and (\ref{eq:sigma_app}) we can then expand $T_\mathrm{t,AI}(E=-e V,\theta)$ to lowest order in powers of $V$ and $\theta$, and obtain
\begin{eqnarray}
T_\mathrm{t,AI}(E=-e V,\theta)&\approx&\frac{\Gamma_\mathrm{L} \Gamma_\mathrm{R}}{\Gamma_\mathrm{L}+\Gamma_\mathrm{R}} \frac{2}{\Delta}\left[1-\frac{3}{2}\frac{e^2 V^2}{\tilde\Delta^2}-\right.\nonumber\\
&&\left.\frac{\pi^2}{2}\frac{k_\mathrm{B}^2\theta^2}{\tilde\Delta^2}
\right],
\label{eq:tvb_app}
\end{eqnarray}
and
\begin{eqnarray}
\frac{\pi^2}{6}\left.\frac{\partial^2 T_\mathrm{t,AI}(E,\theta)}{\partial E^2}\right|_{E=-eV}k_\mathrm{B}^2 \theta^2&\approx&\frac{\Gamma_\mathrm{L} \Gamma_\mathrm{R}}{\Gamma_\mathrm{L}+\Gamma_\mathrm{R}} \frac{2}{\Delta}\nonumber\\
&&\left[-\frac{\pi^2}{2}\frac{k_\mathrm{B}^2\theta^2}{\tilde\Delta^2}
\right],
\label{eq:tvb_app2}
\end{eqnarray}
If we insert these last two equations into Eq.~(\ref{eq:gtheta_app}) we obtain
\begin{equation}
G_0=G_\mathrm{AI}(0,0)=\frac{2 e^2}{h}\frac{4\Gamma_\mathrm{L} \Gamma_\mathrm{R}}{\left(\Gamma_\mathrm{L}+\Gamma_\mathrm{R}\right)^2},
\label{eq:app_g0}
\end{equation}
and
\begin{eqnarray}
\frac{G_\mathrm{AI}(V,\theta)-G_0}{G_0}&\approx&-\frac{3}{2}\frac{e^2 V^2}{\tilde\Delta^2}-\pi^2\frac{k_\mathrm{B}^2\theta^2}{\tilde\Delta^2}
,
\label{eq:g2nd}
\end{eqnarray}
and from this relation we can directly extract the lowest order expansion coefficients of the conductance (see Eq.~(14)
of the main manuscript) as
\begin{eqnarray}
c_V&=&\frac{3}{2},\label{eq:app_cV}\\
c_\theta &=&\pi^2.\label{eq:app_ctheta}
\end{eqnarray}
These results are valid in the strong correlation limit ($U \gg \Gamma$) and at particle-hole symmetry, and agree with the general values obtained in Ref.~\citenum{Nozi.74_esi,Munoz.11_esi} and in Sec. \ref{sec:app_conductancerSPT} for this case. We note that the full energy dependent transmission contains more information than this lowest order expansion, which is valid only in the low voltage region.

\subsection{Approximations for large temperatures}
As voltages or temperatures get large, the second order expansion discussed so far will eventually deviate significantly from the correct result, and we now estimate the magnitude of the voltages and temperatures at which this happens. 

By comparing the exact NRG results for $\Sigma(E,0)$ to the second order expansion (Eq. (\ref{eq:sigma_app})) we find that the terms describing the energy dependence are approximately valid for energies also beyond $\tilde\Delta$, so that they are rather well suited to describe the shape of the Kondo peak at zero temperature. This is also reflected in the fact that the second order expansion of the FWHM of the Kondo Peak agrees well with the exact NRG results (see Fig. \ref{fig:fwhm_nrg}).

\begin{figure}
\includegraphics[width=0.45\textwidth,clip=true]{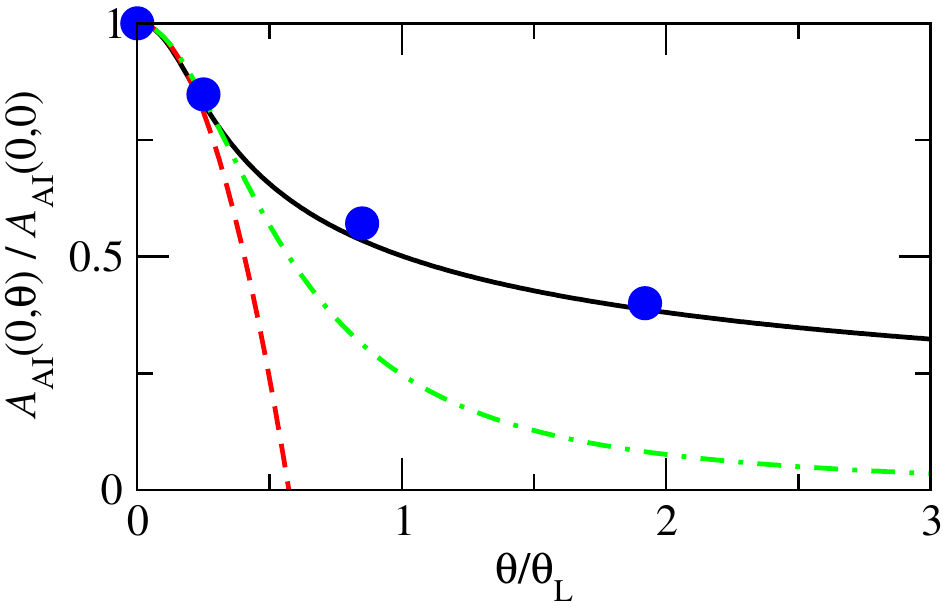}
\caption{Temperature dependent normalized density of states, equivalent to the spectral function, as function of temperature; the NRG results (blue filled circles) are extracted from Fig. 6 in Ref. \citenum{Costi.94c_esi}. The green dash-dotted lines correspond to the relation given in Eq. (\ref{eq:dosmax_app}), the red dashed curve in indicates the exact second order low temperature expansion in Eq. (\ref{eq:app_a2ndTheta}), and the black curve show the DOS fitted to the NRG results using Eq. (\ref{eq:DOSthetaphenom_app}).
} \label{fig:deltafit}
\end{figure}

To separately verify the range of validity of the second order expansion in temperature of the self-energy we evaluate $A_\mathrm{AI}(0,\theta)$ (Eq. (\ref{eq:dose_app})) for increasing temperatures using the second order expansion of $\Sigma(0,\theta)$ given in Eq. (\ref{eq:sigma_app}), and then compare it to the exact NRG results of Ref. \citenum{Costi.94c_esi}. This comparison is presented in Fig. \ref{fig:deltafit}(a), where we plot the normalized spectral function $A_\mathrm{AI}(0,\theta)/A_\mathrm{AI}(0,0)$ as function of temperature, and the green dash-dotted curve is the result of the second order expansion, while the large blue dots are the exact NRG data. It can be seen that the results agree well for low $\theta$, and that even at higher $\theta$ they qualitative behaviour of a decaying height of the Kondo peak is captured by the second order expansion. On a quantitative level however the results start to deviate significantly already at temperatures of about 0.5 $\theta_\mathrm{L}$, with the second order expansion results overestimating the decay.

In analogy of the second order expansion of the conductance presented in the previous subsection (Eq. \ref{eq:g2nd}) we now also expand $A_\mathrm{AI}(0,\theta)$ to second order in $\theta$. Using Eqs. (\ref{eq:dose_app}) and (\ref{eq:sigma_app}) we obtain
\begin{equation}
A_\mathrm{AI}(E=0,\theta)=\frac{1}{\pi \Delta}\left[1-\frac{1}{2}\pi^2 \frac{k_\mathrm{B}^2\theta^2}{\tilde\Delta^2}+O\left(\frac{k_\mathrm{B} \theta^4}{\tilde\Delta^4}\right)\right],
\label{eq:app_a2nd}
\end{equation}
with the second order coefficient $\pi^2/2$ equal to half the size of the second order temperature expansion coefficient for the conductance, $c_T$, given in Eq. (\ref{eq:app_ctheta}). The larger size of $c_T$ is due to the fact that additionally to the change in $A_\mathrm{AI}(E=0,\theta)$ it also captures the temperature induced broadening of the Fermi distribution. Overall we can then expect that the range of validity in terms of magnitude of temperature is similar for $A_\mathrm{AI}(E=0,\theta)$ and for the 0 bias conductance $G(V=0,\theta)$.

We can equivalently express the expansion of $A_\mathrm{AI}(E=0,\theta)$ directly in terms of $k_\mathrm{B}\theta_\mathrm{L}=\frac{\pi}{4}\tilde{\Delta}$
[Eq. (8)
of the main manuscript] as
\begin{equation}
A_\mathrm{AI}(E=0,\theta)=\frac{1}{\pi \Delta}\left[1-\frac{1}{2}\left(\frac{\pi}{2}\right)^4 \frac{\theta^2}{\theta_\mathrm{L}^2}+O\left(\frac{\theta^4}{\theta_\mathrm{L}^4}\right)\right].
\label{eq:app_a2ndTheta}
\end{equation}
The results are plotted as red dashed curve in Fig. \ref{fig:deltafit}. It can be seen that this expansion starts to significantly deviate from the exact NRG results already at about $\theta \approx 0.2 \theta_\mathrm{L}$. Since we expect a similar upper limit of the validity for temperature dependent zero-bias conductance, it shows that to reliably compare experiment to the second order expansion in Eqs. (14)
of the main manuscript, or (\ref{eq:g2nd}),
accurate measurements are needed for temperatures smaller than about $0.2\;\theta_\mathrm{L}$.

In practice it is difficult to accurately measure the conductance below 0.2 $\theta_\mathrm{L}$, and indeed most experiments measure up to temperatures significantly larger than $\theta_\mathrm{L}$.
Therefore, based on the exact NRG results in \cite{Costi.94c_esi} a fitting function for the temperature dependent low bias conductance has been proposed in Ref. \citenum{Goldhaber-Gordon.98_esi}, which has then been found to match well the experimental conductance drop with temperature. The form of this fitting function is
\begin{equation}
G(V=0,\theta)=G_0 \left[1+ a \frac{\theta^2}{\theta_\mathrm{L}^2}\right]^{-s} + G_\mathrm{B},
\label{eq:gvthetaphenom_app}
\end{equation}
with $G_\mathrm{B}$ equal to the background conductance, $G_0=G(V=0,\theta=0)-G_\mathrm{B}$, and with $a$ determined by the imposed condition that $G(0,\theta_\mathrm{L,1/2})$=1/2. Note that $\theta_\mathrm{L,1/2}$ is similar but not identical to $\theta_\mathrm{L}$. For large $U/\Gamma$, as is the case for the Au-PTM system, the value is given as 1.2~$\theta_\mathrm{L}$ in Ref. \citenum{Costi.94c_esi} and as 1.041~$\theta_\mathrm{L}$ in Ref. \citenum{Merker.13_esi}. If we neglect this small difference and assume $\theta_\mathrm{L,1/2}\approx\theta_\mathrm{L}$, then we obtain $a=2^{1/s}-1$. We then determine the value of $s$ from the condition that the second order expansion in $\theta$ of Eq. (\ref{eq:gvthetaphenom_app}) has to correspond to the exact result in Eq. (\ref{eq:g2nd}), which gives the relation $s \left(2^{1/s}-1\right) =\left(\pi/2\right)^4$. This can be solved numerically to give $s \approx0.20$. 

In order to obtain an empirical equation for the temperature dependence of the self-energy we now use an analogous approach to fit NRG results of $A_\mathrm{AI}(E=0,\theta)$ for the whole temperature range shown in Fig. \ref{fig:deltafit}(a). Based on the empirical fitting Eq. (\ref{eq:gvthetaphenom_app}) for the conductance we use a similar functional form to fit $A_\mathrm{AI}(0,\theta)$ to the NRG results:
\begin{equation}
A_\mathrm{AI}(E=0,\theta)=\frac{1}{\pi \Delta}\left[1+ a \frac{\theta^2}{\theta_\mathrm{L}^2} + \left(a \frac{\theta^2}{\theta_\mathrm{L}^2}\right)^2\right]^{-s/2},
\label{eq:DOSthetaphenom_app}
\end{equation}
where $a$ and $s$ have the same values as used in Eq. (\ref{eq:gvthetaphenom_app}). The resulting function is shown as black curve in Fig. \ref{fig:deltafit}(a), and it can be seen that it approximates well the exact NRG results. It is also straight forward to verify that this fitting function has the correct second order expansion given in Eq. (\ref{eq:app_a2ndTheta}).

\section{Expansion coefficients of the rSPT conductance}
\label{sec:app_conductancerSPT}
\begin{figure}
\includegraphics[width=0.45\textwidth,clip=true]{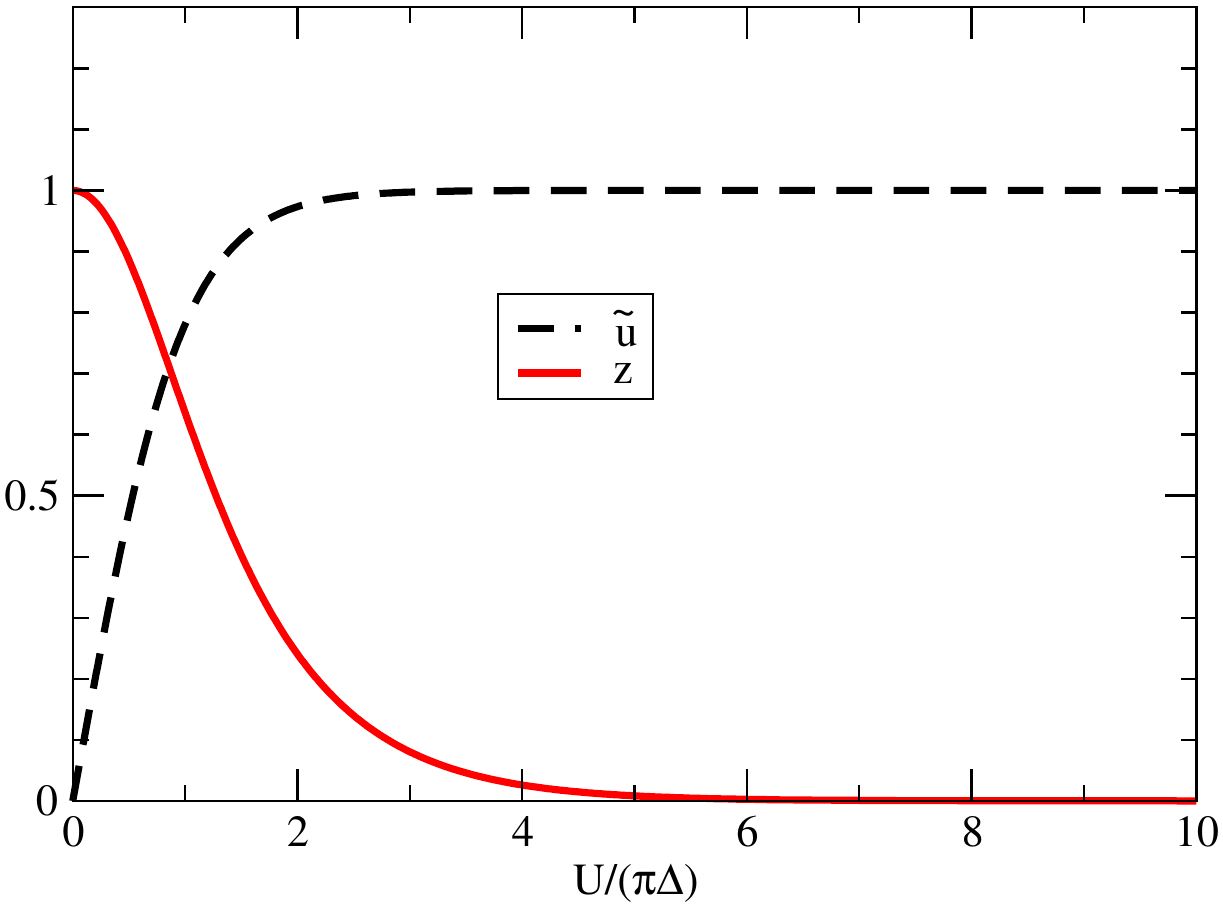}
\caption{The Bethe ansatz results for the wave function renormalization factor, $z$, and the rescaled interaction $\tilde{u}$, as defined in Sec. 5
of the main manuscript. Note that the strong coupling fixed point value $\tilde{u}=1$ is already reached for $U/(\pi\Delta)\approx 2$.
} \label{fig:app_zu}
\end{figure}

In this section, we provide full expressions for all transport coefficients, defined in  Eq. (14)
of the main manuscript. The results are based on the extended version of  the rSPT, introduced in reference \cite{Munoz.16_esi} but have not been presented previously. As a result, the transport coefficients below are correct to higher order in the deviation from the symmetric Anderson model than those in \cite{Munoz.11_esi}. As discussed in section 5
of the main manuscript, the central object of the method is the  wavefunction renormalization factor $z$. For a constant hybridization function, it can be obtained exactly via the Bethe ansatz \cite{Munoz.11_esi}. Fig.  \ref{fig:app_zu} shows $z$ and $\tilde{u}$ obtained from the Bethe ansatz as a function of the bare Coulomb interaction $U$, measured in units of $\pi \Delta$.

For an energy-dependent hybridization function, the NRG can be used to obtain $z$ through $z=(1-\partial \Sigma^R_\sigma(\omega)|_{\omega=0})^{-1}$ as discussed in the main part. Alternatively, $z$ can also be determined perturbatively through
\begin{align}
\label{eq:z-compute}
z&=1+\dfrac{\tilde{U}^2}{2} \int_{-\infty}^{+\infty}\dfrac{d\omega}{2\pi i} \Big[ \tilde{\Pi}^{(0)A}(-\omega)\dfrac{\partial}{\partial \omega}\tilde{g}^{(0)K}(\omega)\nonumber \\
&+\tilde{\Pi}^{(0)K}(-\omega)\dfrac{\partial}{\partial \omega}\tilde{g}^{(0)R}(\omega) \Big] 
\,+O\big(\tilde{U}^3\big),
\end{align}
here $\tilde{U} = z^2 \Gamma_{\uparrow\downarrow}(0,0,0,0)$, where
$\Gamma_{\uparrow\downarrow}(0,0,0,0)$ is the two-particle vertex of the reference system at $\omega = 0, \theta = 0, V=0$.
Then, at the order of approximation used in the underlying expansion, it turns out that $\Gamma_{\uparrow\downarrow}(0,0,0,0) = U + O(U^3)$, and hence up to second order one has $\tilde{U} \sim z^2 U + O(U^3)$; $\tilde{g}^{(0)x}$ is the  renormalized Green function,  and  $\tilde{\Pi}^{(0)x}$ refers to the renormalized polarization bubble of the reference system, {\itshape i.e.}, where $2\epsilon_d+U=0$, indicated by the superscript $(0)$ and  $x=K,A,R$ denotes the Keldysh, advanced, retarded component respectively.
All terms in Eq.~(\ref{eq:z-compute}) have to be evaluated in the zero-temperature ($\theta=0$) limit.
In the case of a constant hybridization function, this equation reduces to 
\begin{align}
\label{eq:const}
2\big(3-\frac{\pi^2}{4}\big)\big(\frac{U}{\pi\Delta}\big)^2\, z\,=\, \sqrt{1+4\big(\frac{U}{\pi\Delta}\big)^2\big(3-\frac{\pi^2}{4}\big)}-1.
\end{align}
For small $U$ we can therefore use this equation to estimate $z$, while for large $U$ we use the $z$ obtained either from NRG or from the Bethe ansatz, as outlined in the main text.

Equipped with $z$ for the particle-hole symmetric SIAM on the Keldysh contour, the method of references \cite{Munoz.11_esi,Munoz.16_esi} yields an approximative expression for the renormalized Green function $G^{x}$ ($x=K,A,R$) of the asymmetric SIAM with energy-dependent hybridization function near the strong-coupling fixed point. Then the imaginary part of $G^{R}$ can be used to extract the spectral function and calculate the current according to Eqs.~\ref{eq:itotal} and \ref{eq:ttai}.
From the resulting conductance $G(V,\theta,B)=dI/dV$, expressions for
the expansion coefficients in Eq. (14)
of the main manuscript can be constructed. 
These are obtained as
\begin{strip}
\par\noindent\rule{\dimexpr(0.5\textwidth-0.5\columnsep-0.4pt)}{0.4pt}%
\rule{0.4pt}{6pt}
\begin{eqnarray}
c_B = \frac{\left(1 - 3\delta^2 \right) \left(1 + \frac{\tilde{u}}{1 + \tilde{\epsilon}_d^2} \right)^2 + \frac{2\delta\tilde{u}\tilde{\epsilon}_d(1 + \delta^2)}{(1 + \tilde{\epsilon}_d^2)^2}}{4\left(1 + \delta^2 \right)^2},
\end{eqnarray}
\begin{eqnarray}
c_{V} = \frac{2 \left(3 \delta ^2-1\right) (\zeta -1)-\left(\delta ^4-1\right) (2 \zeta +1) \tilde{u}^2 + \frac{4 \delta  \left(\delta ^2+1\right) \zeta  \tilde{u} \tilde{\epsilon}_d }{\left(\tilde{\epsilon}_d ^2+1\right)^2}}{2 \left(\delta ^2+1\right)^2},
\end{eqnarray}
\begin{eqnarray}
c_{V E_d} = \frac{2 (1-\kappa ) \delta }{(\kappa +1) \left(\delta ^2+1\right)},
\end{eqnarray}
\begin{eqnarray}
c_{\theta V} &=& \frac{\pi ^2}{6 \left(\delta ^2+1\right)^4} \Bigg \{\left(-12 \left(5 \delta ^4-10 \delta ^2+1\right) (\zeta -1)+\left(\delta ^2+1\right) \tilde{u}^2 \left(-3 \delta ^6 \zeta +3 \delta ^4 \left(\zeta  \left(\frac{4 \tilde{\epsilon}_d ^2}{\left(\tilde{\epsilon}_d ^2+1\right)^4}-7\right)+9\right) \right. \right. \nonumber \\ 
&+&\left. \left. \delta ^2 \left(\zeta  \left(\frac{8 \tilde{\epsilon}_d ^2}{\left(\tilde{\epsilon}_d ^2+1\right)^4}+111\right)-162\right)+\zeta  \left(-\frac{4 \tilde{\epsilon}_d ^2}{\left(\tilde{\epsilon}_d ^2+1\right)^4}-15\right)+27\right)-\frac{24 \delta  \left(\delta ^4-1\right) \tilde{u} \tilde{\epsilon}_d }{\left(\tilde{\epsilon}_d ^2+1\right)^2}\right)\Bigg\},
\end{eqnarray}
\begin{equation}
c_{\theta V E_d} = \frac{2 \pi ^2 (\kappa -1) \left(-3 \delta ^5 \tilde{u}^2 \left(\tilde{\epsilon}_d ^2+1\right)^2+6 \delta ^3 \left(\tilde{u}^2-1\right) \left(\tilde{\epsilon}_d ^2+1\right)^2+3 \delta  \left(3 \tilde{u}^2+2\right) \left(\tilde{\epsilon}_d ^2+1\right)^2+3 \delta ^4 \tilde{u} \tilde{\epsilon}_d +2 \delta ^2 \tilde{u} \tilde{\epsilon}_d -\tilde{u} \tilde{\epsilon}_d \right)}{3 (\kappa +1) \left(\delta ^2+1\right)^3 \left(\tilde{\epsilon}_d ^2+1\right)^2},\\
\end{equation}\\
\vspace{\belowdisplayskip}\hfill\rule[-6pt]{0.4pt}{6.4pt}%
\rule{\dimexpr(0.5\textwidth-0.5\columnsep-1pt)}{0.4pt}
\end{strip}
\begin{eqnarray}
\delta(\tilde{\epsilon}_d) = \tilde{\epsilon}_d - \tilde{u}\taninv \tilde{\epsilon}_d,
\end{eqnarray}
together with the parameters
\begin{eqnarray}
\kappa &=& \frac{\Gamma_{L}}{\Gamma_{R}},\nonumber\\
\zeta &=& \frac{3\kappa}{(1+\kappa)^2}.
\end{eqnarray}
Finally, for $c_\theta$ we obtain
\begin{eqnarray}
c_{\theta} = \frac{\pi^2}{3}\frac{1 + 2\tilde{u}^2(1-\tilde{\delta}^2) + \frac{2\tilde{u}\tilde{\delta}^2}{(1 + \tilde{\epsilon}_d^2)^2} - \frac{4\tilde{\delta}^2}{1 + \tilde{\delta}^2}}{1 + \tilde{\delta}^2},
\end{eqnarray}
where the quantity $\tilde{\delta}$ appearing in this expression is given by
\begin{eqnarray}
\tilde{\delta}=\tilde{\delta}(\tilde{\epsilon}_d,\tilde{u}) = \tilde{\epsilon}_d - \dfrac{\tilde{u}}{1+\tilde{u} }\taninv \tilde{\epsilon}_d.
\end{eqnarray}

We note that for particle-hole symmetry ($\tilde{\epsilon}_d=0$), strong coupling ($\tilde u \approx 1$) and $\Gamma_L\ll\Gamma_R$ ($\kappa\approx 0$), we recover the results of Eqs. (\ref{eq:app_cV}) and (\ref{eq:app_ctheta}), namely $c_\theta=\pi^2$ and $c_V=3/2$.
Note also that an expansion of the coefficients to lowest order in $(1-\kappa)$, which corresponds to an expansion around the symmetric coupling case ($\Gamma_L=\Gamma_R$), shows that in this case the 0$^\mathrm{th}$ order term is non-zero only in $c_{V}$ and $c_{\theta V}$, the first order term is non-zero only in $c_{V E_d}$ and $c_{\theta V E_d}$, and the second order term is again non-zero only in $c_{V}$ and $c_{\theta V}$. This odd-even behaviour in the expansions is a consequence of the more general relation $G(\theta,V)=G(\theta,-V)$ valid for $\kappa=1$, and which combined with Eq. (14) of the main manuscript requires that $c_{V}$ and $c_{\theta V}$ can only have even powers of $(1-\kappa)$, while $c_{V E_d}$ and $c_{\theta V E_d}$ can only have odd powers of $(1-\kappa)$.




\providecommand*{\mcitethebibliography}{\thebibliography}
\csname @ifundefined\endcsname{endmcitethebibliography}
{\let\endmcitethebibliography\endthebibliography}{}

\end{document}